\documentstyle[12pt,epsf,aaspp4]{article}
\newfont{\rmlil}{cmr5}
\newcounter{curtabno}
\newcounter{curfigno}
\makeatletter
\def\tabnum#1{\def\thetable{#1}\let\@currentlabel\thetable
    \addtocounter{table}{\m@ne}}
\def\fignum#1{\def\thefigure{#1}\let\@currentlabel\thefigure
    \addtocounter{figure}{\m@ne}}
\def\raggedright{\let\\=\@centercr
    \@rightskip\z@ plus .3\hsize
    \rightskip\@rightskip
    \leftskip\z@skip
    \pretolerance=1
    \hyphenpenalty=1 \exhyphenpenalty=1
    \parindent\z@}
\def\leavemode{\unhbox\voidb@x}
\def\@cite#1#2{{#1\if@tempswa , #2\fi}}
\def\@biblabel#1{}
\newenvironment{tableletters}{\refstepcounter{table}%
    \setcounter{curtabno}{\value{table}}%
    \let\@curthetab=\thetable%
    \edef\cur@tab{\csname thetable\endcsname}%
    \def\thetable{\arabic{curtabno}\Alph{table}}%
    \setcounter{table}{0}}%
    {\let\thetable=\@curthetab%
    \setcounter{table}{\value{curtabno}}}
\newenvironment{figureletters}{\refstepcounter{figure}%
    \setcounter{curfigno}{\value{figure}}%
    \let\@curthefig=\thefigure%
    \edef\cur@fig{\csname thefigure\endcsname}%
    \def\thefigure{\cur@fig\Alph{figure}}%
    \setcounter{figure}{0}}%
    {\let\thefigure=\@curthefig%
    \setcounter{figure}{\value{curfigno}}}
\makeatother
\newenvironment{centertable} %
   {\begin{table}[htbp] \begin{center}} %
   {\end{center} \end{table}}
\newenvironment{minipagetable}[1] %
   {\begin{table}[htbp] \begin{center} \begin{minipage}{#1} %
    }%
   {\vspace{-0.1in} \end{minipage} \end{center} \end{table}}
\newenvironment{centerfigure} %
     {\begin{figure}[htbp] \begin{center}} %
    {\end{center} \end{figure}}
\newcommand{\cste}{Cs$_2$Te}
\newcommand{\csi}{CsI}
\newcommand{\mgftwo}{MgF$_2$}
\newcommand{\caftwo}{CaF$_2$}
\newcommand{\srftwo}{SrF$_2$}
\newcommand{\siotwo}{SiO$_2$}

\newcommand{\hii}{H\small{II}\normalsize}

\newcommand{\dexp}[1]{$\times 10^{#1}$}

\newcommand{\flx}{erg cm$^{-2}$ pix$^{-1}$ \AA$^{-1}$ s$^{-1}$}

\newcommand{\surfbr}{erg cm$^{-2}$ \AA$^{-1}$ s$^{-1}$}
\newcommand{\eunitps}{E-unit pix$^{-1}$ s$^{-1}$}
\newcommand{\um}{$\mu$m}
 
\newcommand{\ea}{et~al.}
\newcommand{\eg}{e.g.}
\newcommand{\ie}{i.e.}
\newcommand{\cf}{cf.}
\newcommand{\Astro}[1]{{\em Astro-#1}}
\newcommand{\AstroMiss}{{\em Astro}}
\newcommand{\UIT}{UIT}
\newcommand{\HUT}{HUT}
\newcommand{\WUPPE}{WUPPE}
\newcommand{\ANS}{{\em ANS}}
\newcommand{\IUE}{{\em IUE}}
\newcommand{\OAO}{{\em OAO-2}}
\newcommand{\GHRS}{GHRS}
\newcommand{\patent}{\raise0.55ex\hbox{\kern0.101ex$^{\hbox{\rmlil %
R}\kern-1.251ex\bigcirc}$\kern0.101ex}}
\newcommand{\idl}{IDL}
\newcommand{\twodash}{-{}-}
\newcommand{\threedash}{-{}-{}-}
\newcommand{\reffigure}[1]{Fig.~\ref{#1}}
\newcommand{\showsect}[1]{Sec.~\ref{#1}}

\newcommand{\Strut}{\rule{0pt}{2.25ex}}
\newcommand{\SStrut}{\rule{0pt}{0.2ex}}
\newcommand{\TabNewLine}{\tabularnewline}
\newcommand{\TabLine}{\hline}
\newcommand{\DTabLine}{\TabLine\omit\SStrut\TabNewLine\TabLine}
\newcommand{\parcaption}[2]{\parbox{#1}{\caption{#2}}}
\newcommand{\centerepsf}[1]{{\centering \leavemode \epsffile{#1}}}

\lefthead{Stecher, \ea}
\righthead{\UIT\ Instrument and Data Characteristics}
\begin{document}
%
%
\title{The Ultraviolet Imaging Telescope: Instrument and Data Characteristics}
\author{
Theodore P. Stecher\altaffilmark{1},
Robert H. Cornett\altaffilmark{1,2},
Michael R. Greason\altaffilmark{1,2},
Wayne B. Landsman\altaffilmark{1,2},
Jesse K. Hill\altaffilmark{1,2},
Robert S. Hill\altaffilmark{1,2},
Ralph C. Bohlin\altaffilmark{3},
Peter C. Chen\altaffilmark{1,4},
Nicholas R. Collins\altaffilmark{1,2},
Michael N. Fanelli\altaffilmark{1,2},
Joan I. Hollis\altaffilmark{1,2},
Susan G. Neff\altaffilmark{1},
Robert W. O'Connell\altaffilmark{5},
Joel D. Offenberg\altaffilmark{1,2},
Ronald A. Parise\altaffilmark{1,4},
Joel Wm. Parker\altaffilmark{6},
Morton S. Roberts\altaffilmark{7},
Andrew M. Smith\altaffilmark{1},
and
William H. Waller\altaffilmark{1,2}
}
\altaffiltext{1}{Laboratory for Astronomy and Solar Physics, NASA/GSFC, Code
680, Greenbelt, MD 20771}
\altaffiltext{2}{Hughes STX Corporation, 4400 Forbes Boulevard, Lanham, MD
20706}
\altaffiltext{3}{Space Telescope Science Institute, Homewood Campus, Baltimore,
MD 21218}
\altaffiltext{4}{Computer Sciences Corporation, NASA/GSFC, Code 684.9, 
Greenbelt, MD  20771}
\altaffiltext{5}{University of Virginia, P.O. Box 3818, Charlottesville, VA
22903}
\altaffiltext{6}{Department of Space Science, Southwest Research Institute,
Suite 426, 1050 Walnut Street, Boulder, CO  80302}
\altaffiltext{7}{National Radio Astronomy Observatory, Edgemont Road,
Charlottesville, VA  22903}
%
%
\clearpage
\begin{abstract}
The Ultraviolet Imaging Telescope (\UIT) was flown as part of the
\AstroMiss\ observatory on the Space Shuttle Columbia in December 1990 and
again on the  Space Shuttle Endeavor in March 1995.   Ultraviolet
(1200-3300\AA) images of a  variety of astronomical objects, with a 40\arcmin\ 
field of view and a resolution of about  3\arcsec, were recorded on 
photographic film.  The data recorded during the first flight are available to
the astronomical community through the National Space Science Data Center
(NSSDC); the data recorded during the second flight will soon be available
as well.  This paper discusses in detail the design,  operation, data
reduction, and calibration of \UIT, providing the user of the data with 
information for understanding and using the data. It also provides
guidelines for analyzing other astronomical imagery made with image
intensifiers and photographic film.
\end{abstract}
%
%
\clearpage
\section{Introduction and UIT Overview}\label{uitoverview}

The Ultraviolet Imaging Telescope (\UIT) is the only astronomical 
telescope that can produce images of faint,  ultraviolet-emitting objects 
with a resolution of 3\arcsec\ over a 40\arcmin\ field of view (FOV).  \UIT\ 
images emission in the range 1200--3300~\AA\ through broadband filters and a 
grating, enabling the study of diverse astronomical targets (\eg, Galactic 
nebulae, globular star clusters, nearby galaxies, and clusters of galaxies).
As part of the \AstroMiss\ payload complement of three co-mounted 
ultraviolet instruments, \UIT\ flew on Space Shuttle missions in December 1990
and March 1995.  This paper is intended to provide users of \UIT\ data with 
the information and understanding needed to use the data. It presents detailed 
specifications for the technical aspects of \UIT's  performance, describes how 
they were derived, and discusses their implications for the scientific use of 
the data.

\UIT\ is an f/9 Ritchey-Chr\'{e}tien telescope with a 38~cm aperture.  It 
contains two selectable detector systems, or cameras, each of which is a 
magnetically focused,  two-stage image intensifier with a  phosphor output
that is fiber-optically  coupled to 70~mm film. The  far-ultraviolet (FUV, or
``B'') camera's intensifier has  a \csi\ photocathode,  while the near-UV (NUV,
or ``A'') camera has a \cste\ photocathode.   The cameras' two film transports
each carry enough film for about $\sim 1200$ exposures, which are digitized to
produce image arrays.  Each camera  has a six-position  filter wheel; the NUV 
camera can be used with a diffraction  grating for low-dispersion full-field
spectroscopy.  The  \UIT\ photocathode-filter combinations yield excellent
long-wave rejection  with negligible ``red leak'' for most
applications.  An articulated  secondary mirror  provides internal image motion
compensation.  \reffigure{uitcutaway}\ shows a cutaway view  of \UIT.
Table~\ref{uitspec}\ gives general specifications for \UIT\ and its data; 
\cite{blum} and \cite{stecher} provide more details on the hardware.  

\UIT\ images are recorded on Kodak IIa-O film and are  digitized
with Perkin-Elmer 1010m microdensitometers.  The digital
images are processed by software called the Batch Data Reduction (BDR)
system, which  linearizes and flat fields them.  BDR also computes
astrometric parameters, performs aperture and Point Spread Function (PSF)-fit 
stellar photometry (based upon DAOPHOT [\cite{stetson}]),  rotates 
images to north-up orientation, and corrects for image-tube distortion.  
Processed data is
archived in the  Flexible Image Transport System (FITS; \cite{wells},
\cite{fitsgsfc}), which is the standard format for astronomical images and
data tables (\cite{hilla}).

During the \Astro{1}\ mission (1990 December 2--10),  \UIT\ took 361 near-UV
and 460 far-UV exposures of 66 targets.  During the \Astro{2}\ mission
(1995 March 2--18), \UIT\ took 758 far-UV exposures of 193 targets.   
The \cste\ camera failed at the launch of  \Astro{2}\
and did not operate on orbit; therefore, no near-UV astronomical data were
acquired during that mission.

\subsection{\UIT\ Scientific Objectives and Results}\label{uitscience}

\UIT\ was designed to make wide-field, high-resolution, solar-blind
ultraviolet images.  The 40 arcmin field, in particular, was  selected to
image most globular clusters, galaxies, and  clusters of galaxies in single
pointings.  Because \UIT's bandpass is until now nearly unobserved at high
resolution, \UIT\ images have provided a wealth of discoveries and 
possiblities for new analysis.  \UIT\ scientific results include studies of 
Galactic reflection nebulae (\cite{witt}), supernova remnants (\cf\
\cite{cornett}), and many globular clusters (\cf\ \cite{landsmanc}); 
analysis of supernova SN1987A as well as several Magellanic Cloud fields 
(\cf\ \cite{hillc}); detailed analysis of the stellar content,  \hii\
regions, and large-scale structure of nearby spiral  (\cf\ \cite{jhilla}),
and dwarf galaxies (\cite{jhillb}); analysis of the  UV spatial and
photometric properties of the old population in galaxies  (\cf\
\cite{oconnell}); observations of clusters of  galaxies including detailed
studies of remarkable members  (\cf\ \cite{esmitha}); a large-scale study
of the UV sky (\cite{waller});  and a near-UV bright object catalog
(\cite{esmithb}). Initial \UIT\ scientific results from the \Astro{1}\ flight
were presented in  the 1992 August 10 issue of {\em The Astrophysical
Journal (Letters)}. An up-to-date list of \UIT\ publications is maintained on
the \UIT\ Home Page at {\em http://fondue.gsfc.nasa.gov/UIT\_HomePage.html}. 

Two representative \UIT\ images are shown in \reffigure{mplate}\ and 
\reffigure{ngcplate}.  

\reffigure{mplate} compares a \UIT\ 639-second A1 image
($\lambda_{eff}=2490$\AA) of the spiral galaxy M81 with a ground-based
visible-band image made at Kitt  Peak National Observatory.  The UV image
accentuates the UV-bright Population I component in the spiral arms, as
well as the hot old-population stars in  the nuclear bulge. The weakness of
UV emission from the general disk  population in this early-type spiral is
typical (\cite{jhillc}); later type  spirals show stronger UV flux from the
disk (\cite{cornetta}). 

\reffigure{ngcplate} is a 781-second exposure of the globular cluster
NGC 6752 made in the B5 filter ($\lambda_{eff}=1620$\AA).  The effective
rejection of visible-band light suppresses light from the cluster's 
$\sim$~100,000 main sequence stars, leaving only the 355 hot, UV-bright 
horizontal branch stars in the field of view.  The \UIT\ image resolves hot
stars in the cluster core, and the 40 arcminute field of view encompasses 
the entire cluster, providing the first complete census of hot horizontal 
branch stars in NGC 6752 (\cite{landsmand}). The overexposed bright object 
is a foreground star.  

\subsection{Optics and Image Motion Compensation}\label{optimcinf}

The \UIT\ telescope optics are of conventional Ritchey-Chr\'{e}tien design
with \mgftwo\ coatings.  \reffigure{suboptical}\ shows  the suboptical
assembly, which moves selected filters and the grating into the optical path. 
A rotating diagonal mirror directs converging  light from the
secondary mirror to either of the two detector systems, and an additional
fixed mirror in each detector light path permits more compact packaging.  A
six-position filter wheel is positioned $\sim 70$~mm in front  of each 
detector.  The filter complement includes uncoated plates of fused quartz,
crystalline quartz, \mgftwo, and \srftwo, as well as metal-dielectric
interference filters on substrates of fused quartz, crystalline quartz, and
\caftwo.  The highest-throughput filters are the uncoated plates; the plate
defines the short-wavelength limit while the cathode's photoelectric
threshold defines the long-wavelength limit. Parfocalization assures that
each of the two sets of filters has a common focal  plane.  The
plane-parallel transmission filters introduce axial and lateral chromatic
aberrations and some spherical aberration in the converging light  beams,
which are partially corrected by plano/convex lenses in each of the
\mgftwo\ entrance windows of the detectors.  Details on the filters  are
presented in \showsect{bandpass}\ and the wavelength response curves for
each filter-detector combination are shown in \reffigure{nuvcurves}\ and
\reffigure{fuvcurves}.  A transmission grating (system spectral resolution 
19~\AA) can be moved into the NUV camera 
optical path. 

Internal baffling assures a minimum of two scatterings before light from
outside the FOV reaches the detector.  An external, cylindrical sun baffle
mounted ahead of the telescope contains annular sub-baffles and a polished
cone, which deflects any light that is incident at angles
greater than $51^\circ$ from the telescope axis.

An Image Motion Compensation (IMC) system stabilizes the \UIT\ image to a
finer tolerance than is possible using only the Instrument Pointing System
(IPS) of the Spacelab platform, which permits a few arcseconds of payload
wobble.  The IMC system's two gyroscopes and a CCD star tracker provide 
attitude update information at 50 Hz.  These digital inputs are fed to 
an 18-Hz-bandwidth closed-loop analog system that tilts the \UIT\ secondary 
mirror through small angles about two axes, counteracting the payload motion 
to minimize image motion in the focal plane.  Measurements of inflight IMC 
performance are discussed in \showsect{focnpsf}

\subsection{Image Intensifiers, Electronics, and Film}\label{iielectfilm}

The \UIT\ detector system comprises two high-quality image intensifiers
which down-convert each incoming ultraviolet photon into many blue
photons.   The image intensifiers are focused with samarium-cobalt magnets which
produce a uniform 165~gauss magnetic field parallel to the axial electric 
field in the  intensifiers, which causes the photoelectrons to follow 
one-turn helical paths when accelerated through 13~kV. Field uniformity is 
$\sim 1$\%.  The high-voltage power supplies are located in the film
vaults,  which are maintained at a pressure of 1~atm to prevent coronal
discharge and film dehydration. The  cathodes are at
ground, and the high-voltage anodes are at 26~kV.  A potential of either
$+700$~volts or $-100$~volts is applied to a fine mesh grid near each
cathode to serve as an electronic shutter.  

Each of the photocathode-window combinations  provides good UV response and is
insensitive to long-wavelength light; the cutoff for \cste\ is at
$3500$\AA, and the cutoff for \csi\ is at $2000$\AA.  Green light
output by the first image-intensifier stage in each detector is generated
by a P20 phosphor screen, which is coupled through fiber optics to the
bialkali photocathode of the second stage.  Blue light output by the
second stage is generated by a P11 phosphor screen, which is coupled  to
the film by  fiber optics.  A conductive carbon backing on the film
prevents the accumulation of a static charge. 

Annotation devices controlled by the Dedicated Experiment Processor (DEP;
see  \showsect{depflight}) project the frame count, exposure time, a
30-step gray scale and fiducial marks onto each frame.  The exposure time
is written as a four digit number $nnne$, where the time is $nnn.\times
10^e$ $\mu$s.  Each film  transport holds about 1200 frames.  The film is
advanced by stepper motors controlled by the DEP.  A platen holds the film
against the fiber optic bundle and is retracted by a solenoid for frame
advances.

The Kodak IIa-O film used for calibration and both flights is from a  
single batch purchased in 1984 and stored frozen since  then.  Flight film
was transported  in dry ice  to Kennedy Space Center (KSC), where it was
loaded into \UIT\ ten to  fourteen days before launch.  Film temperature,
digitally monitored and recorded, was
controlled and held to cool room temperatures (18-25~C) during flight, recovery,
and transporation back to Goddard Space Flight Center (GSFC)),  except 
for a 2-day interval of cooling which reached $-2$~C at the end of \Astro{2}.  

\subsection{\UIT\ Computer and Flight Operations}\label{depflight}

The DEP controls the operations of the \UIT\
and its interaction with the onboard and ground-based observers and
Spacelab. It consists of two identical Motorola M6800 microprocessor
systems with memory, input/output ports, and special-purpose hardware (\eg,
timers).  

All \UIT\ operations during \Astro{1}\ were planned to be conducted on board the
Shuttle by a Payload Specialist (PS) and a Mission Specialist  using
two Data Display Units (DDUs; \ie, computer terminals) together with a
joystick controller for manual guiding.  Acquisition images from the
Hopkins Ultraviolet Telescope (\HUT) and Wisconsin Ultraviolet
Photo-Polarimeter Experiment (\WUPPE) television cameras were to be provided 
to the PS. 
However, the failure of both DDUs 3 days into the mission required a new
procedure.  In this operational mode, the \AstroMiss\  instruments were
commanded from the Payload Operations Control Center (POCC) at the Marshall
Space Flight Center (MSFC).  The Mission Control Center at  Johnson
Space Center generated IPS commands, while the Alternate Payload
Specialist  in the POCC examined \HUT\ acquisition images on a
television monitor and talked the on-board PS through target acquisition
and verification.  The PS guided the telescope manually with the
joystick.  

During \Astro{2}\ the flight crew usually conducted routine configuration
that applied to all three instruments (\eg, acquisition, pointing, and
other IPS operations).  The ground crew at the MSFC POCC commanded the
individual instruments when practical.  In particular, \UIT\ personnel edited
and initiated exposure sequences to ensure the use of the proper
filters on either side of the orbital day/night boundary.  When the ground
crew could not perform these operations (due to loss of signal from the
orbiter or because high priority ground commanding was being done
elsewhere), the on-board crew performed them.

\subsection{Film Development and Digitization}\label{filmdevdig}

The film was developed at GSFC in a Houston Fearless automated film
processor.  The processor first removes the film's conductive carbon
backing and then develops the film for 240 seconds in Kodak D-19, which is
followed by conventional stop, fixing, and wash  baths.  When the
\Astro{1}\ film was developed, an additional wash bath was used to assure
the removal of all the carbon backing; this was deemed unnecessary for
\Astro{2}.

Photographically recorded data from \UIT\ are digitized using two identical
Perkin-Elmer Model 1010m microdensitometers at GSFC.  Fiducial marks
exposed on the film  with the images are used to align each frame with a
repeatability of 20~\um\ or  better thereby ensuring consistent
registration.    All \Astro{1}\ frames were scanned  with a 20~\um\ square
aperture and a sample spacing of 20~\um\ (corresponding to $\sim 1.12$ arcsec 
on the film).  In addition, some portions of some frames were also scanned 
with a 10~\um\ ($\sim 0.56$ arcsec) spacing (termed ``sub-stepped''), affording 
better resolution and a marginally better  signal-to-noise (S/N) ratio.  All
\Astro{2}\ frames were scanned  with a 20~\um\ square aperture and a sample
spacing of 10~\um, and the high-resolution density images were retained;
these digitized images were boxaveraged to an effective sample spacing of
20~\um\ for further processing.  Background fog levels were measured from
patches of unexposed film at the edges of the frame, surrounding the 
exposed sky image.  Microdensitometer drift  in gain was removed by fitting
a bilinear surface to  these fog measurements and subtracting this surface
from the digitized output (see \showsect{dignfog}).

A simple naming standard indicating camera, mission, and frame number 
identifies each digital image.   Each image name is of the form
``CuvMnnn''.   `C' is `N' for the near-UV camera or `F' for the far-UV
camera. The value of `M' identifies the frame count series: it is is `0' for 
\Astro{1}; since a DEP memory change reset the frame counter during \Astro{2}, 
'M' is `1' for frames taken before the DEP memory switch, and `2' for frames 
taken after the DEP memory switch.  `nnn' is the three digit frame number.  
For example, the 69$^{th}$ \Astro{2}\ far-UV frame following the DEP memory 
switch is designated ``FUV2069.''

\subsection{Data Reduction and Analysis}\label{dataredintro}

\UIT\ standard data products are produced using the BDR system at GSFC (see
\showsect{dataproduct}).   BDR is written in Fortran 77 and C, and consists
of  several steps whose application is controlled by FITS header
parameters. The latest version of  BDR runs on a DEC 3000 (Alpha)
workstation.  Working storage is  magnetic disk, while the data is archived
on Exabyte tape and on the UniTree data storage facility of the NASA Center
for Computational Sciences (NCCS), with additional deep backups. 
The current calibration data and the BDR source code is
available through the  National Space Science Data Center (NSSDC).  The
software is provided solely for documentation,  with no support for outside
installation or use.

Interactive processing and analysis of \UIT\ data is done with the
Multi-Option \UIT\ Software System Environment (MOUSSE) system.  MOUSSE is
written in  \idl, a commercial\footnote{Research Systems, Inc., 2995
Wilderness Place, Suite 203, Boulder, CO  80301} programming, plotting,
and  image display language available for most Unix workstations as well as
for OpenVMS, Microsoft Windows, and other platforms.  \idl\ is well adapted
to interactive work and is easy for users to extend.  As a result,  MOUSSE
is a loosely controlled, continuously growing software system.  Many MOUSSE
routines are available through the NSSDC (like BDR, without support);
MOUSSE routines are also available through the IDL Astronomy User's Library
(\cite{landsmanb}).

%
%
\section{Laboratory and Flight Calibration Procedures and Results}
\label{labfltcal}

\UIT\ calibration data were obtained at GSFC in December
1984 through February  1985, when focus, bandpass, flat field, linearity, and
absolute calibration procedures were performed.  Following each flight,  
science  data were used to rederive   the linearity and absolute
calibration results.  Flight images of populous star fields were used to 
develop models of geometrical distortion (due to the magneto-optical nature 
of the detectors; see \showsect{geomdist})  of the images.

Several cycles of calibration and reduction, each based on the previous
cycles results, were performed on \Astro{1}\
data.  The final version of \Astro{1}\ data
processing is known as the FLIGHT15 data stream, and the final version for 
\Astro{2}\ data is called FLIGHT22. 

\subsection{Focus and Point Spread Function}\label{focnpsf}

\UIT\ focusing was performed using a mercury  light source  and a collimator 
in air in a light tunnel at GSFC.  An initial rough  mechanical focus was 
followed by electronic detector focusing through adjustment of the image 
intensifier voltages.  Mechanical focus was then repeated for several 
locations across the field of view.  Electronic focus was repeated  after 
installation of the telescope in its outer structure in order to compensate 
for changes in the magnetic environment of the detectors.  A final focus 
check was performed in vacuum.

Following each mission, \UIT\ focus and PSF characteristics were
determined from measurements of large numbers of stellar images in 
flight data frames  of rich star fields distributed throughout the mission  
timelines.  For the entire fields of images showing no evidence of 
pointing problems (such as elongated PSFs), the average FWHM of a 
two-dimensional Gaussian fit to NUV stellar images is 2.92\arcsec; to 
FUV images, 3.36\arcsec.   Coma is apparently present at large field radii, as
evidenced by elongation of stellar images there in the field-radial direction.
However, for radii smaller than 80\% of the full field, coma is negligible. 
For those radii the FWHM values are 2.71 and 3.03\arcsec\ for NUV and FUV 
\Astro{1}\ 
images, respectively.  These values represent the typical quality of the PSF in 
\Astro{1}\ and \Astro{2} \UIT\ data taken under nominal conditions. 
All measurements described above were performed on the  digitized images, and 
therefore include the effects of convolution of the  PSF with the 20~\um\ 
microdensitometer aperture, which adds 0.4--0.5\arcsec. See 
Table~\ref{uitpsfdesc} for a summary of PSF measurements.

There are about two dozen contributors to \UIT\ PSF size.  The
image intensifiers have a specification of 40 line pairs mm$^{-1}$ at 
the Hg 2527\AA\ line, corresponding to a FWHM contribution of $\sim 1.7$~arcsec.
The actual contribution at shorter wavelengths is larger because
higher-energy photons cause electrons to leave the first photocathode 
with non-zero kinetic energy, leading to defocussing.  Therefore the image 
intensifier contributions are an important part of the PSF size.  Other 
important contributors include: defocussing due to chromatic aberration in 
quartz and \mgftwo\ windows and filters; telescope collimation and focus 
errors; spherical aberration caused 
by converging beams impinging on the planar filters; uncorrected focal 
surface curvature; and finite digitization aperture.  For most exposures, 
however, stellar images over the center part of the field are symmetric and 
similar to one another,  with the nominal sizes listed in 
Table~\ref{uitpsfdesc}. 

Pointing difficulties, evidenced by asymmetric stellar images over much of
the field,  dominate the measured PSF in about a third of \UIT\
images.   Perhaps the worst case is presented by the \Astro{2}\ images 
FUV2695 through FUV2897; the mean FWHM 
for 7 frames in this interval is $4.53 \pm 0.39$\arcsec, and the PSFs are 
systematically elongated in one direction. Unfortunately, the observer's log 
contains no indication that there were any problems during these 
observations; the cause has not been identified.   During other observations, 
large excursions in the IMC system were observed to be correlated with crew 
motion on board the orbiter.

See \reffigure{psfsurf}\  for a surface plot of 
a nominal \UIT\ PSF generated from stellar images on a single frame. 
\reffigure{diffpsf}\ shows a cross-section through the measured PSF of 
\reffigure{psfsurf}\, the best-fit Gaussian for that PSF, and the
difference between the measured and fit PSFs.  It shows 
that the \UIT\ PSF is sharper at its core than a true Gaussian.

\subsection{Bandpass}\label{bandpass}

Bandpass characterization consists of measuring the response of \UIT\ as a 
function of wavelength for each filter and the grating.  These procedures 
were performed at GSFC in February 1985.  The total response of \UIT\ was
determined for each of 11 filter bandpasses and for the grating (A6) mode. 
The complete instrument, except for its outer skin, was  placed in a vacuum
tank.  \UIT\ viewed a light source consisting  of an auxiliary Cassegrain
telescope equipped with a scanning monochromator with a hydrogen lamp at
its focus.  Two photomultipliers, calibrated against National 
Institute for Standards and Technology (then
National Burea of Standards)  standard photodiodes, were placed at the 
\UIT\ image intensifier output
and  in the beam of the auxiliary telescope, respectively.  As the
monochromator  scanned each  filter bandpass, the ratio of the two
photomultiplier outputs provided the  relative response of the \UIT\ with
wavelength.   See Table~\ref{uitfilters} for a description of all filters; 
see \reffigure{nuvcurves}\ and \reffigure{fuvcurves}\ for the filter
response curves.

\subsection{Grating}\label{grating}

In addition to UV filter imagery, \UIT\ provides low-dispersion UV spectra
across the field of view using a transmission grating.

The grating is ruled at 75 lines mm$^{-1}$ on a \caftwo\ substrate with a
center thickness of 4.920 mm.  The blaze angle is $1.16^\circ$ and the
linear dispersion is  840 \AA\ mm$^{-1}$ in the first order. Spectra are
recorded in the range  $1400 \leq \lambda \leq 3400$~\AA\  with a
resolution of $\sim 19$~\AA\ over a $\sim 35^{\prime}$ FOV.  The spectral
resolution is limited by the image tube detector rather than by the optics.
The ruled lines of the grating are slanted with respect to the edges of
the  film, producing spectra angled $45.73^\circ$ from the film travel
direction.   For most sources, the zero-order image and first-order
spectrum are recorded; however, for bright sources, the  second order is
also detected.  

Grating images are initially processed in the same manner as other \UIT\
fields. To obtain spectra, the digitized image is rotated so that the
spectra extend in the $\pm$Y direction; then the extraction is done using
a numerical slit with a user-selectable length (in the $\pm$X direction).  
An interactive software package for reducing and analyzing grating spectra
was written in  \idl\ and made available as part of the MOUSSE package
through the NSSDC.

The grating output was calibrated in the laboratory with the same lamp, 
monochromator and auxiliary telescope as were used for the bandpass 
characterizations.  However, film was used as a detector rather than a 
photomultiplier, thus producing a laboratory absolute flux calibration 
for the grating.  A post-flight absolute flux calibration was derived using 
the observed \UIT\ spectrum of G191B2B, a white dwarf standard (\cite{bohlin}). 
which was compared to IUE library spectra (see \reffigure{gratecurv}).

\UIT\ obtained grating spectra in 8 fields during the \Astro{1}\ mission.  Of
this set, 6 frames had sufficient exposure time to record $\sim 20 - 30$
spectra per frame.  The faintest source for which a useful spectrum could be
detected in the  longest exposures had a UV magnitude of $\sim 16.5$ in the A1
filter.  From the spectrum of QSO~1821+643, the
limiting flux in a 1000~s exposure was $\sim 1 \times 10^{-15}$ \surfbr\ at
2000~\AA.

Spectral features due to Fe~II $\lambda 2600$, Mg~II $\lambda 2800$ and the
$\lambda 2200$ interstellar absorption are easily seen in the \Astro{1}\ data.
Classification of extracted spectra is found to be possible by using the
library of UV stellar spectra from \cite{fanellia} smoothed to the spectral
resolution of the grating data.  \UIT\ appears well-suited for  studies of
the variation of the 2200~\AA\ feature.  See \cite{fanellib} for a more
detailed description of the grating observations and results.  

\subsection{Flat Fields}\label{flatfields}

\UIT\ flat field images of a surface illuminated uniformly by broad-band UV
radiation are used to remove
variations in sensitivity as a function of position in the FOV from astronomical
images.  Flat fields were taken both in the laboratory and during
\Astro{1}.  However, the ones made in flight, which were of the bright
earth, were spoiled by afterglow from the output phosphor of the image
tubes, and were not used.  Some daylight sky images taken during
\Astro{1}\ were approximately flat fields.  These images were analyzed as
part of a study of UV sky brightness, but were not used as calibration
data. 

Laboratory flat field exposures were made with a set-up
similar to that used for bandpass characterization.  The hydrogen lamp was
used without the monochromator.  A finely ground, aluminized quartz  flat
was positioned to obscure the secondary mirror of the auxiliary telescope 
and to serve  as a Lambert surface.  For each camera-filter combination, 4
sets of 5 exposures each were made, except for filter 2 of the B camera, 
which required prohibitively long exposure times.  The five images in each
set covered a range of exposure times, of which the smallest and the
largest differed by a factor of sixteen. The film was processed and 
digitized following the same procedures as used with the flight film, as 
described in \showsect{filmdevdig}.

From each five-image sequence, the two best-exposed flat fields were selected, 
yielding eight examples for each filter (of which a few were spoiled by 
development mishaps).  In addition, the data set taken to determine the 
characteristic curve also consists of flat fields, and several of them
were  included.  The resulting collection of data is the basis for
detailed  studies of \UIT\ densitometry and flat field reproducibility, and 
for creating the final \UIT\ flat fields  themselves.  

Composite flat fields were made for processing flight data. One such flat field
was computed for each filter-camera combination from a set of 8--12
well-exposed laboratory flats selected as described above.  A median
combination was used to stack the input images.   However, permanently flawed 
pixels (\eg, those caused by \UIT\ photocathode or output phosphor defects)
were left at their original values; this procedure minimized the  possibility
of confusing end users of the data, since science images  contained the same
flaws.  The composite flat field derived for the ``B5'' filter is shown in
\reffigure{flatfield}.

The flat fields were re-processed for FLIGHT22 reduction.  Several small 
defects (see Table~\ref{ffdefects}\ for locations) had been smoothed in 
\Astro{1} processing but, as they existed in both science and flat field 
images, were reinstated into the flat fields for \Astro{2}. A 1-pixel
misalignment  between the \Astro{2}\ data  and the flat fields was
also discovered during this analysis and corrected.  In addition, flat fields
for images with \Astro{2}'s 10-micron digitization step size were produced 
by bilinear interpolation of the original flat fields.

The quality of flat fielding at high resolution is limited by the 
uncertainty in aligning frames with one another.  \UIT\ frames are aligned 
on the microdensitometer
stage by  placing 2 fiducial marks on the film at established locations. 
After setting the first fiducial to its prescribed coordinates, the
operator rotates the stage to bring the second fiducial to its prescribed
coordinates. The alignment was verified experimentally by measuring the
position of the second fiducial on a set of 27 (digital) images of 8 film frames
which were  repeatedly set up in the standard way and digitized.  
The resulting measured uncertainty is  $\sigma = 0.95$~pixel (rms of $x$ and 
$y$ uncertainties).    Therefore, typical pixel alignment uncertainty for 
\UIT\ images is $\sim 1$~pixel.  This justifies the addition of flat field 
images to make composite flat fields as well as the doubling of flat field 
size by interpolation for substepped images. 

The photometric quality of the flat fielding procedure was also tested over 
large image scales.  Mean flux values for $128 \times 128$-pixel  ($143 \times
143$ arcsec) regions over the entire image circle agree within 2.65\%
(1-$\sigma$), implying that spatial variations due to uncertainties in 
flat fielding are less than that value.  

\subsection{Characteristic Curve}\label{charcurv}

The \UIT\ characteristic curve is a function that maps the  photographic
density in each pixel to the relative UV energy incident on the telescope at
that image location.  Application of the characteristic curve to the digitized
density images linearizes \UIT\ data.  In practice, the \UIT\ characteristic
curve is a 4096-element look-up table, which is the same for both
cameras.   The linearized data frames are arrays of integers that are
proportional to the time-integrated flux detected in each pixel for a given
exposure.  The units of these pixels are termed {\em E-units\/}.  A 
multiplier, supplied in the header of each image,  converts E-units to  
physical units.

Laboratory characteristic-curve data are actually flat fields of varying
densities, taken with the same hardware setup as the flat fields intended
for correcting spatial variations in sensitivity, but over a much larger
range of exposure times.  For each camera, two sequences of fourteen
exposures each  were obtained with successive exposure times
in a 2:1  ratio, so that the total dynamic range of each sequence is
$\sim 16000$.   A trial characteristic curve
was generated by an iterative algorithm that  linearizes the image
sequences, forcing the output E-unit values to  increase by factors of 2 in
step with the exposure time.  

This procedure may also be applied to non-flatfield images, such 
as flight images of stellar fields, by comparing corresponding pixels of
different exposures, binned by E-unit value.   Almost every \UIT\   
observation supplies characteristic-curve data, since normal sequences for 
each filter comprise 2 or 3 frames with exposure times in particular ratios 
(\eg, 5:25:1 for most \Astro{1}\ observations, and 1:10 for \Astro{2}).  
These data, as well as more extensive sequences from planned calibration 
observations, were used to test and rederive the \UIT\ characteristic curve.  

Analysis leading to the FLIGHT15 characteristic curve determined that the 
curve initially derived from laboratory  data and the curve derived from 
stellar  photometric data were significantly different.  Because flight-acquired
stellar data were much more abundant and more relevant to observations,  
a new characteristic curve was derived based primarily on \Astro{1}\ 
photometry of stars.  This curve proved to be as good for extended-source
flight data ({\em e.g.} from the reflection nebula NGC 7023) as the original 
flat field curve, and was therefore adopted for FLIGHT15.  

Stellar flight data were also used to determine the FLIGHT22 characteristic 
curve. For the FLIGHT22 reduction, the parameters of a characteristic curve 
with a  specific mathematical form were determined from stellar photometry 
measurements of multiple images of the same region of the sky.  For each 
iteration, the density images are converted to exposure (E-units) using 
the following expression:  

\begin{eqnarray*} \mbox{log}_{10} E & = & \left[ a_0 +
\frac{a_1}{1 + \omega} + \frac{a_5}{(1 + \omega)^2} \right] + \\ &   & a_2 (
\omega - \omega_0 ) + a_3 ( \omega - \omega_0 )^2 + a_4 ( \omega - \omega_0
)^3 \\ \omega   & = & 10 ^ {D / 820} - 1 \\ \omega_0 & = & 10 ^ {200 / 820} -
1 \\ 
\end{eqnarray*} 

\noindent (\cite{owaki}), where $D$ is the measured density and $\log_{10}
E$ is the  exposure in E-units.  The linearized counts for objects in each
frame are compared to the exposure times of those frames. The coefficients
in the above expression are adjusted to force the linearized  count ratios
to match the  time ratios using a grid-search fitting technique  as
described in \cite{bevington}.   The algorithm begins with the laboratory
characteristic curve and  adjusts the curve in successive steps constrained
by the flight data photometry.  For FLIGHT15 only the upper end of
the curve was adjusted, while the lower  end (the toe) was held equal to
the preflight curve, because the lower end of the curve proved to be 
difficult to improve with the parametrization used.   The use of the 
\cite{owaki} parametrization of the curve in FLIGHT22 made it possible to 
fit the entire curve to photometry data.  See \reffigure{charcurvplot}\ for  
both curves.

This technique for deriving a characteristic curve (forcing linearized pixel  
values from different images of the same source to have the ratio of the 
images' exposure times) assumes that reciprocity failure is negligible. 
For \UIT\ this is expected because the large photon gain ($\sim 400$) of the 
image intensifiers amplifies each detected photon enough to create a 
developable grain in the emulsion.  The fact that a stable curve can be 
derived for flight frames, with their wide range of source flux levels, 
confirms that this is true.

\subsection{Noise Characteristics}\label{noisechar}

\UIT\ noise characteristics are determined from repeated measurements of
the  pixel values in laboratory flat fields; characteristic curve data are
the  primary source of these measurements.  Each characteristic  curve frame
was flat fielded, and data from corresponding parts of each of the 
resulting images  were compared.  Because these images are flat fielded,
each pixel is an independent measurement of a single value---a
fixed number of E-units.  The variation in the measurement of this value, 
as averaged over large and small angular scale sizes, measures the  \UIT\
noise characteristics at a single location in the FOV.  The resulting \UIT\
noise/signal ratio is  small (15--20\% for single pixels and 5--7\% for
regions several arcseconds in size) over the center of the dynamic range,
with substantial increases at the extremes.  Uncertainty in 
the flat fields adds up to 3\% for comparison of sources in different field 
locations. See \reffigure{noisecurvplot}\ for the \Astro{1}\ and \Astro{2}\
noise functions.

It is important to note that noise increases rapidly at the low signal 
levels where sky and faint-source surface brightness are often measured.  
In addition, the fundamental film and microdensitometer fog level uncertainty 
(see Section~\ref{dignfog}), corresponding to about 1 E-unit, is an important 
contributor to the uncertainty in absolute measurements at low surface 
brightnesses.  See Table~\ref{uituncert} for an overview.

\subsection{Microdensitometry and Fog Level}\label{dignfog}

Images are scanned using Perkin-Elmer Model 1010m microdensitometers, which
produce image arrays in  ``PDS density units'' ($P$; PDS
is a nickname for a microdensitometer), related to conventional
photographic density, $D = \log (1/T)$, by $P \approx D/800$.  On this
scale, \UIT\ film densities range from  $P \sim 320$ at fog level to $\sim
3000$ at film saturation.  A constant standardizing shift  moves this scale
down by $P\approx120$ (to $200 \rightarrow \sim 2880$) before 
conversion to linear units.  This forces the  mean fog level to $P \sim
200$ which the characteristic curve sets to an exposure level of zero,
allowing the fog level to be ignored during later analysis.

An important part of the digitization process is to establish the fog level
of the film, which corresponds to the 0 E-unit level.  This is
accomplished  by adjusting the gain and offset values of the
microdensitometer photomultiplier electronics to fixed values at an opening
in the film (\ie, a sprocket hole), effectively ``zeroing'' the
microdensitometer.  This procedure  determines the fog level density at the
beginning of the scan.   

To ensure data repeatability during scanning, a step wedge was scanned 
daily, and the scan compared to fiducial examples.   These comparisons show 
how the photometric response of the microdensitometer changes over time. 
Typical results show that microdensitometer 
density measurements averaged over a few dozen  pixels agree to 1 part in 
300 through the density range of IIa-O film, implying  that the 
microdensitometer contribution to \UIT\ noise is negligible. 

Fog levels were measured within the \UIT\ film frame and outside the
circular area containing astronomical data. Inspection of flight-data scans
showed that the background fog levels at these zero-exposure levels differed 
from one another within each frame, and that the fog level varied
smoothly from one position to another.  As measured at the
microdensitometer, the mean fog   value for \UIT\ images was $P=323$ PDS
density units, with the corners typically  varying from the mean by 3
units.  The resulting absolute uncertainty in flux measurements, corresponding
to an uncertain additive constant for large areas on the frame, is
estimated to be 1 E-unit near zero flux.  It is not clear whether these 
variations are caused by microdensitometer drift, by optical effects (for 
example thin-film effects arising from the air layer between the film and 
the microdensitometer glass),  or by true spatial variations in unexposed  
film density.  

BDR removes background level variations  and ``flattens'' the fog value by
removing a two-dimensional background using a bilinear interpolation 
technique to model the fog across each image from mean densities
determined for each of the four image corners.  At this stage images are 
also adjusted to a standard fog value of 200 density units by subtracting a 
constant.

\subsection{Absolute Calibration}\label{abscalib}

Absolute calibration of \UIT\ was estimated before flight by comparing 
laboratory data with previously obtained sounding-rocket data,  in
order to confirm the performance of the instrument.  This absolute
calibration was  verified and refined by comparing \UIT-measured fluxes 
with results from \IUE, \OAO, \HUT, \ANS, \GHRS, and other
UV instruments. The primary calibration basis is common observation of
stars by \IUE\ and \UIT.

\UIT\ images containing the stars used for calibration were processed with 
an \idl\ version of BDR using the FLIGHT15 characteristic curve,
flat fields, and  corrected exposure times.  The flux of each star was
measured using aperture  photometry extending to aperture radii large
enough that  the growth curve is flat (typically 14-20~pixels).  The \IUE\ 
spectra used the Finley absolute calibration (\cite{finley}),  which is
similar to the latest NEWSIPS calibration (\cite{nichols}). In addition, 3
spectra obtained by \HUT\ during \Astro{1}\ were also used.   Each
measurement was assigned errors based upon \IUE\ and \UIT\ photometry which
were used to compute a weighted mean calibration constant for each camera. 

The resulting values of $C_{NUV}$ and $C_{FUV}$, which are the constants of
proportionality between physical units and E-units, are:

\begin{center}
\begin{tabular}{rl}
\multicolumn{2}{c}{\Astro{1}\ / FLIGHT15} \TabNewLine
$C_{NUV}$ &= 1.25\dexp{-16} $\frac{\hbox{\flx}}{\hbox{\eunitps}}$\TabNewLine
$C_{FUV}$ &= 6.15\dexp{-16} $\frac{\hbox{\flx}}{\hbox{\eunitps}}$\TabNewLine
&\TabNewLine
\multicolumn{2}{c}{\Astro{2}\ / FLIGHT22} \TabNewLine
$C_{FUV}$ &= 6.88\dexp{-16} $\frac{\hbox{\flx}}{\hbox{\eunitps}}$\TabNewLine
\end{tabular}
\end{center}

\noindent  where $C_{NUV}$ was based on 23 measurements of 13 \IUE\
stars, \Astro{1}\ $C_{FUV}$ was based on 43 measurements of 18 \IUE\ 
stars,  and the \Astro{2}\ $C_{FUV}$ was based on 75 measurements of 49
\IUE\ stars.  These constants apply to 20-\um\ pixels of 20-\um\ separation.
Since the number E-units is proportional to the number of pixels, constants
for images with other pixel separations must be scaled to pixel density per
unit sky solid angle.  In practice, absolute calibration is applied through 
`BSCALE', a multiplicative constant within each image  header, that converts 
the integer E-unit values to flux in \flx.  Typical uncertainties in 
absolute fluxes for \UIT\ sources were determined  to be $\sim 15$\%.

The \Astro{2}\ calibration for FLIGHT22 was determined in the same fashion,
using the FLIGHT22 characteristic curve, flat fields, and corrected exposure
times.  The basic FLIGHT22 calibration constant derived was 1.12 times the to
FLIGHT21 calibration constant.  

However, an important anomaly was discovered during this analysis.  The 
absolute calibration computed using \IUE\ spectra decreases with the 
exposure time of each image, so that \UIT\ appears less sensitive for 
longer exposure times.  Although the effect is reminiscent of reciprocity 
failure, it is not--it depends on exposure time, not source brightness. 
   
The physical origin of this anomaly is not 
understood.  However, a satisfactory empirical correction is made (and
incorporated in the processed BDR data products) by multiplying the nominal 
absolute calibration by the exposure time in seconds raised to the 0.1 
power (see \reffigure{abscalfig}).  The correction is effective for sources 
of all brightnesses on a given frame (a range of up to $\sim 500$). 
For FLIGHT22 data products, the 
exposure time correction factor is stored in the FITS keyword `TIMEFAC',  
but users should still use the FITS header variable `BSCALE' to convert 
images to physical units (\ie, `BSCALE' incorporates the `TIMEFAC' correction).
No such anomaly was found in \Astro{1}\ data, and no corrections are made for 
it. Users are encouraged to calibrate each individual image 
independently whenever possible, with data from other UV instruments.   
  
The calibration was also checked for extended sources.  For example, 
measurements of the reflection nebula NGC7023 in
\Astro{1}\ and \Astro{2}\ images were made, and compared to \IUE\ spectra.
The \UIT\ measurements are approximately 15\% brighter than the \IUE\ 
measurements;  given the errors inherent in the \UIT\ and \IUE\
calibrations and the uncertainty in \IUE\ positions,
the difference in the measurements is reasonable.

\subsection{Geometrical Distortions}\label{geomdist}

The \UIT\ image intensifiers introduce small but significant geometrical
image distortions ranging in size from a few
arcseconds near the field center to  20--30\arcsec\ at the edge. The 
distortions have the form and the same origin as the classic S-distortion 
commonly found in image tubes, although the amplitude is small.  

It was determined from \Astro{1}\ data that the image distortion is a fixed
pattern for all measured frames, and that it was possible to produce \UIT\
images that were rectified to the precision of their resolution by applying 
a single
correction for each camera.  This approach, which was adopted, permits
a rectified image to be generated with no accurate knowledge of the 
astrometry of the frame (often the case in the relatively uncharted UV sky), 
and effectively separates the process of geometrical correction from 
astrometric solutions.
 
Image distortions  were removed by employing a third-order polynomial 
transformation, fitted by comparing \UIT\ images of rich star fields to 
digitized Guide Star Survey plates.  It has the form:

\begin{eqnarray*}
x' &=& a_0 + a_1 x + a_2 y + a_3 x^2 + a_4 xy + a_5 y^2 + a_6 x^3 +
a_7 x^2 y + a_8 x y ^2 + a_9 y^3\\
y' &=& b_0 + b_1 x + b_2 y + b_3 x^2 + b_4 xy + b_5 y^2 + b_6 x^3 +
b_7 x^2 y + b_8 x y ^2 + b_9 y^3\\
\end{eqnarray*}

\noindent A separate solution, or distortion model,
was derived for each camera, and was used to reduce the distortions to $\leq
3$\arcsec\  (\cite{greason} ), comparable to the PSF size.  
Distortion models were computed from both \Astro{1}\ and \Astro{2}\ data;
Table~\ref{nuvimgdistmodelA}, Table~\ref{fuvimgdistmodelA}, and 
Table~\ref{fuvimgdistmodelB} show the coefficients.
The two models differ by several arcseconds near the image edge. The
\Astro{1}\  model was used for reductions before and including FLIGHT21.
The \Astro{2}\  model was used for the FLIGHT22 reduction.  The \Astro{2}\
model is the better determined model, since the available positional data
is more uniformly distributed over the frame than was the case for
\Astro{1}.

The distortion correction is incorporated into astrometric solutions by 
transforming the image positions of the known objects into ``distortion-free'' 
space before computing the plate solution. The correction is also applied 
to the images at the
processing stage in which the images are resampled and transformed to
standard north-up, east-left orientation.  This step produces images with
orthogonal RA and Dec axes and uniform plate  scales.  These images may be
overlaid onto ones made at other wavelengths, and register perfectly 
within uncertainties due to the PSF size.  The plate solutions in
these images are standard and do not depend upon the distortion model.

For \UIT\ data products, the presence of the geometric distortion model in an
image plate solution is  indicated by the FITS keywords `CTYPE1 =
RA{\threedash}UIT', `CTYPE2 = DEC{\twodash}UIT' for the \Astro{1}\ model,
and `CTYPE1 = RA{\twodash}UIT2', `CTYPE2 = DEC-UIT2' for the \Astro{2}\
model.  The MOUSSE astrometry software  recognizes these non-standard FITS
keywords and applies a distortion correction before computing celestial
positions.  Standard astronomical data systems, such as IRAF, do not
recognize these keywords and will not compute celestial coordinates
correctly.  However, the  north-up, distortion-corrected versions of the
\UIT\ images have the standard tangent-plane geometry (`CTYPE1 =
RA{\threedash}TAN',  `CTYPE2 = DEC{\twodash}TAN') recognized by most
astronomy software systems.

\subsection{Astrometric Solutions}\label{platesoln}

Astrometric solutions are based on the Hubble Space Telescope Guide Star
Catalogue (GSC; \cite{lasker}).   Use of the GSC results 
in plate solutions for the J2000.0 equinox and the FK5 reference frame.  
Because stars of
later spectral type than  mid-A emit very little UV flux, only a minority
of frames show enough GSC stars for a direct solution, \ie, deep exposures
in the NUV. \Astro{1}\ solutions for deep FUV frames were usually
bootstrapped from the NUV, using ultraviolet sources in common between the
two bandpasses.  For many frames, neither of these strategies was adequate,
and others were employed, as follows:  

\begin{enumerate}
\item 
A shallow exposure was bootstrapped from a deeper one of the
same object in the same camera (not necessarily the same acquisition).
\item
The solution for a shallow exposure was copied outright from a
deeper one, provided it was in the same acquisition and the same camera.
\item
The solution of an FUV frame was transformed using standard
coefficients from the solution of an NUV frame in the same acquisition 
(\Astro{1}\ only).
\item
In especially sparse fields,
a rough solution was derived by hand from star and galaxy coordinates in
SIMBAD.
\item
For \Astro{2}\, it was found that ``IPS coordinates'' (RA, Dec, and roll
supplied by the IPS resolvers) produced a solution accurate to an arcminute
or so.   
\item
Where no legitimate sources were identified, no plate solution was
obtained and the astrometric parameters were left in terms of pixels.

\end{enumerate}

Under the best conditions, method 1 or 2 was used with interactive 
software on a workstation.  This solution was either refined by BDR using a
simple  algorithm to match guide stars (from GSC or elsewhere) to
ultraviolet  sources detected on the frame, or it was accepted outright.
The best solutions have $\sim 40$ matches and errors of 2\arcsec\ or less
in both $x$ and $y$.  In cases where this algorithm could not be used, and
one of the alternative types of solution was required, errors were
estimated by the operator.  All plate solutions include the effects of
image distortion.   The measured $x$ and $y$ coordinates of detected UV
sources are corrected to a distortion-free system before a standard
six-parameter solution is computed.  The six parameters are the RA and Dec
of the optical axis and the four partial derivatives of the standard
coordinates $\xi$ and $\eta$ with respect to $x$ and $y$.  

\subsection{Summary of \UIT\ Measurement Uncertainties}
\nopagebreak
\UIT\ instrumental uncertainties are summarized in Table~\ref{uituncert}.

%
%
\section{\UIT-Specific Data Anomalies}

\subsection{Exposure Times}\label{anomexptime}

During the analysis that led to the FLIGHT15 datastreams and data products,
it was determined that exposure times annotated on the film by the \UIT\ DEP
during \Astro{1}\ were incorrect by known amounts.  A DEP software patch which
corrects this problem was written for use during \Astro{2}\ but, due to 
other flight software problems, could not be used.  Therefore, corrected 
exposure times were generated and recorded in the FITS headers as the 
value `EXPTIME', and 
are used in all subsequent processing.  

The uncertainty in the resulting exposure time values is complicated but
not likely to be significant. For the  vast majority of exposures, in  
which there is no IMC shutter gating ($>$90\%), the uncertainty is of order 
0.04~sec  and has no effect on scientific results.  

\subsection{Scratches}\label{anomscratch}

Although clean-room procedures and other careful precautions were
exercised,  the processing and handling of the \UIT\ \Astro{1}\ film caused
a  significant number of longitudinal (film travel direction) scratches
apparently due to contamination of the film by dust.  The scratches in 
general are not at repeatable positions.
The number and size of the scratches has increased as the film has aged,
so  that, in general, frames produced by later scans had more scratches
than  those scanned earlier.    The scratches cause both cosmetic and
scientific  difficulties including uncertainties in the photometry
of particular  sources contaminated by them, and a positive
bias in sky measurements.

Software was written to detect and remove scratches. Typically, between 150
and 600 scratches per image, averaging 50 $\pm$40 pixels in length, were
detected and removed.  The scratch detection algorithm finds candidate
scratches by comparing each column to a pair of comparison columns  on 
either side of it, with criteria for minimum scratch length and depth. 
After all scratch detection is done, scratches
are removed by linear interpolation between the comparison columns.

The scratch-removal software was run routinely in FLIGHT15 processing just 
before the geometric transformation stage, when the image was  put into a
north-up, east-left  orientation.  Therefore, all \Astro{1}\ ``north-up
undistorted intensity'' images (see  \showsect{dataproduct}), boxaveraged
or not, were descratched.

Several improvements in film processing and digitizing procedures, as well
as redesigned film handling fixtures, were employed on the \Astro{2}\ film,
essentially eliminating scratches.  Therefore scratch removal was 
unnecessary for \Astro{2}. 

\subsection{Turn-on Spots}\label{anomspots}

During the first few minutes after the high voltage is turned on, the image
tubes exhibit regions of decreased sensitivity at fixed locations in 
each camera, with typical sizes of 10--40 pixels in radius.  They are 
generally seen only images with very high ($\sim 20-30$~E-unit) backgrounds
such as those caused by scattered sunlight in NUV camera images.  A good 
example is the image NUV0324 of the daytime UX-UMa field (\Astro{1}),
where the central few pixels of the turn-on spots can have zero measured  
flux, in contrast with the sky's value of $\sim 24$~E-units 
(1.6~\dexp{-16}~\flx).  Roughly half of each mission's orbits included a 
passage through the South Atlantic Anomaly (SAA), a region of anomalously 
high charged particle density. To reduce the possibility of damage to the 
cameras and to avoid high fog levels in the film, \UIT's high voltage power 
supplies were turned off during passage through the SAA.  Therefore, turn-on 
spots are a possibility  in images taken throughout both missions.

There are more turn-on spots in the NUV camera than in the FUV camera;
however, the FUV  camera has a particularly troublesome one about 1\arcmin\
from the \Astro{1}\ IPS pointing center. The turn-on spots are also present
in \Astro{2}\ data  (\eg, see image FUV2026 of the Cygnus Loop), although
their presence is  not as striking due to the generally lower sky
backgrounds during the \Astro{2}\ mission.  The turn-on spots have been 
catalogued, and the procedure DARKSPOT was written in MOUSSE to
display them.  Although turn-on spots are easily detected only in the 
presence of a high background, they can affect stellar photometry even in 
images with a very low background.  

\subsection{Film Stripes}\label{anomfilmstripe}

Many \Astro{2}\ images have a wide, faint stripe running vertically down
the center.  The stripe is up to 3 or 4~E-units in magnitude, and has the
effect of raising the background level in the stripe.  It varies in
position, width, shape, and magnitude.  The stripe is on the film and
therefore is not an  effect of digitization; beyond that, the cause is not
certain.  It was not seen in \Astro{1}\ data.  The stripe does not affect 
measurements which use ``local'' sky
measurements such as stellar photometry of the DAOPHOT type.  Large-scale
sky measurements must be carefully performed on striped images.  For 
cosmetic purposes, an interactive MOUSSE routine, STRIPEX, produces
``de-striped'' images.

\subsection{``Measles''}\label{anommeasle}

We use the term ``Measles'' to denote the bright spots seen only in \UIT\ images
of   objects with extremely high visible-light fluxes, such as Jupiter, 
$\alpha$ Orionis, and Luna (the Moon).  Measles are small  (40--60 \um\
across), isolated, and fairly evenly spread over the images  (400--600 \um\
apart).  They exhibit varying intensities that are higher  than the
background.   While measles are not a concern in virtually all the \UIT\
images, they are a major concern in the analysis of lunar data, both
cosmetically and photometrically.

The measles are probably produced by visible light passing through pinholes
in either the  output phosphor of the first stage or the bialkali
photocathode of the  second stage of the \UIT\ \csi\ image tube. In effect,
this light is a red  leak that redistributes light from visually bright
sources over the image, in addition to the detected UV light.  Because the
second stage is  $\sim 8$ cm from the focal plane, the measles produced by
a point source should occur in an annulus a few pixels wide with
a radius of  roughly 200 pixels, centered on the source.

Two rough estimates were made of the magnitude of the problem.  For
Jupiter, the flux in the ``measle'' image was 6\% of the B5 filter's flux
and was distributed in a ring of radius $\sim 170$ pixels around the
planetary image. From scaled Mars images, a magnitude $V=2.7$ was required
to produce a mean flux of 0.1 E-units in 1000 seconds over the annulus
affected.  The effect is therefore insignificant except for solar
system objects and for stars like $\alpha$ Orionis, which are extremely
red, yet so bright that \UIT\ could still observe them.  

The phenomenon is not yet thoroughly investigated. It is not known, for
example, whether the effect scales with exposure time.  It is clear  that
flux from measles is not ultraviolet; therefore, several techniques have
been developed to remove measles from images.

%
%
\section{\UIT\ Data Products}\label{dataproduct}

Standard \UIT\ data products are images at several stages of processing and
with various effective resolutions and tables of aperture and PSF-fit
point-source photometry.    FITS format is used for all products.  Besides
the standard keywords, the header contains information about the frame
and its exposure, the camera, the filter, the  target, the equinox of
observation, and the pointing and orientation of the instrument.  Full
descriptions of the data products themselves are in  \cite{hilla}.

Image products include ``raw density'' images, ``intensity'' images, 
``box-averaged intensity'' images, ``north-up undistorted intensity'' 
images, and ``box-averaged north-up undistorted intensity'' images.  As 
described in \showsect{filmdevdig}, selected \Astro{1}\ frames are
digitized with a smaller  step size than standard (10~\um\ instead of
20~\um), resulting in ``substepped'' images.  All \Astro{2}\ frames are
digitized with a 10~\um\ step size.  Table~\ref{dataprodsumm} summarizes
the standard image products and the \UIT\ filenaming conventions for these
products.

``Raw density'' images are the digitized microdensitometer output with a 
FITS header attached.  All images are arrays of 16-bit integers; 
microdensitometer output  density values range from about 0 to 4096.  The
raw fog value is  at approximately 300 units and film saturation occurs at
about 3000.  For \Astro{1}\, standard density images are  produced by scans
with 20~\um\ pixels and 20~\um\ spacing and are 2048$\times$2048 pixels;
for \Astro{2}\, the pixel spacing is 10~\um\ and the images, which cover 
the same 40.96 mm square region on the film, are 4096$\times$4096 pixels.  
In \Astro{2}\ processing, the images are boxaveraged to 2048$\times$2048
pixels  after linearization but before flat fielding.

``Intensity'' images are photometrically and  astrometrically calibrated. 
They are linearized and flat fielded, and have plate solutions (in
some cases rough ones).   Astrometric solutions are derived and placed
in the header in the form of UIT-specific (e.g. `CTYPE1 = RA{\threedash}UIT', 
`CTYPE2 = DEC{\twodash}UIT') FITS astrometry parameters which are recognized 
and used by MOUSSE and BDR software.  The astrometric solutions are corrected 
for the effects of image distortion, so that high-quality ($\sim 3$\arcsec) 
RA and Dec positions for objects can be calculated by UIT software; but 
distortion is still present in the images themselves.  Since these images have 
not been resampled, they are the preferred type for most photometry.  The 
image arrays contain 16-bit integers with values proportional to integrated 
intensity (intensity $\times$ exposure  time); valid data is in the range from  
$\sim -25$ to 32767.  Negative pixel values reflect noise at fog 
(zero-exposure) level and enforce a satisfactory distribution of values near 
sky; pixels with a value of 32767 are saturated.  The header parameter 
`BSCALE' is the multiplier which converts the image integer values to physical 
flux units.

``Box-averaged intensity'' images are intensity images boxaveraged by
factors of 4$\times$4, reducing their array size to 512$\times$512 for 
convenience in processing and display.  Astrometric and photometric
calibration  constants are adjusted to the new image scale.

``North-up undistorted intensity'' images are 2048$\times$2048 intensity
images  resampled and rotated to north-up, east-left format with the
distortion removed.  The astrometry parameters are transformed
accordingly.  These images have residual distortions  which are less than
3\arcsec\ (hence ``undistorted''), ensuring that they  align with the UV
sky, as well as with other data, to the limit of \UIT\ resolution. In
FLIGHT15 processing, these images are also descratched as described above. 
\Astro{2}\ data are not descratched.

``Box-averaged north-up undistorted intensity'' images are 4$\times$4 
boxaveraged versions of ``north-up undistorted intensity'' images.  These
images are ``descratched'' only if their parent images are   (\ie, in
FLIGHT15).  

``Substepped'' images may be produced to retain the 10~\um\ pixel spacing of 
some \Astro{1}\ scanned images and all \Astro{2}\ scanned images, but are not
a standard data product. All have standard  20~\um\ (1.12\arcsec) pixels, so that standard calibration values
are correct, and the smaller spacing yields improved resolution and
slightly improved S/N.  A few images from \Astro{1}---\ie, the  deepest
exposures of selected fields---have been processed in this way.  Substepped 
image processing capability is present only in the interactive MOUSSE system;
substepped versions of \UIT\ frames are not generally available.

In addition to image products, tables of point source photometric 
measurements are produced for all frames.  For selected crowded frames 
point source fitting is used.  All photometry is performed using an
adaptation for \UIT\ of the DAOPHOT package  (\cite{stetson}),  with the
error analysis modified to account for the \UIT\ detectors, which differ
greatly from CCDs in their noise characteristics.   

The photometric
measurements reported in these point-source tables are in magnitudes where
(\cite{stecher}):

$$\mbox{\it mag} = -2.5 \mbox{log}_{10} (\mbox{\it Flux}) - 21.1.$$

Aperture photometry is derived from ``intensity'' images. Point-source
detection uses a modified version of the DAOPHOT FIND  algorithm.  Aperture
photometry of the detected sources is done with a  modified version of the
DAOPHOT APER algorithm.  The parameters used are  described in chapter 6 of
\cite{hilla}.  The photometry tables  themselves are standard FITS ASCII
format, and they contain positions, data and error  estimates for all
detected sources.  Photometry apertures of 2, 3, and 7  pixels radius are
employed. 

PSF-fit photometry is also computed for some frames using modified DAOPHOT
procedures.  For each frame, a PSF is computed from several isolated,
well-exposed stars as the sum of a gaussian and residuals.  Photometry is
done by scaling the PSF, within a small fitting radius, to match the
detected point sources, with simultaneous fitting of closely spaced stars. 
This technique permits more accurate photometry of crowded sources than
aperture photometry. PSF-fit tables are similar to aperture photometry
tables.  PSF photometry is adjusted for consistency with aperture
photometry of isolated sources using a 3-pixel radius.

Stellar photometry 
performed by BDR uses the ``local sky'' tech\-nique of
its DAO\-PHOT ancestor, with \UIT-specific statistical methods for determining
the sky value. The sky value is computed from an annulus with
inner radius $\sim 15$~pixels and  outer radius $\sim 25$~pixels,
concentric with  the stellar aperture of
3-pixel radius. BDR chooses the mode or the mean to compute the  output sky
level, depending on the measured sky values.  For typical small sky  values
near the toe of the characteristic curve, the mean is used. The photometric 
error specified in BDR photometry
output includes sky error, often as a significant part. Useful \UIT\ 
photometry is obtained on stars with peak pixel E-unit values as low as
the 20's and total fluxes $\sim 200-300$, for apertures or PSF-fit
equivalents with areas of $\sim 30$ pixels.  

All the \Astro{1}\ data products discussed above are available to the
public through the NSSDC.  The \Astro{2}\ data products will soon be
available as well. The flat fields and the \Astro{1}\ and \Astro{2}\
characteristic curves are also available through the NSSDC.

%
%
\acknowledgments

We would like to thank the following people for their assistance in
producing, reducing, and calibrating the \UIT\ data.  Laura Allen assisted
in digitizing the \Astro{2}\ flight film; Kathy Rhode digitized much of the
laboratory calibration data.  Ray Ohl compared the \UIT\ calibrations with
\IUE\ for extended sources.  David Alexander, Kanav Bhagat, Joel Hegg,
Melissa Marsh,  Melanie Menella, Jennifer Picket, and Brendan Smith
performed many of the measurements used to calibrate the \Astro{1}\ and
\Astro{2}\ \UIT\ data.

Funding for the \UIT\ project has been through the Spacelab Office at NASA
headquarters under Project number 440-51.
%
%
%

%
%
\begin{centertable}
\parcaption{2.1in}{\UIT\ Specifications\label{uitspec}}
\begin{tabular}{lp{3in}}	
\Strut\TabNewLine
\DTabLine
\Strut Clear Aperture		   &   38 cm			 \TabNewLine 
\Strut Primary Mirror Focal Length  &  144.211 cm		  \TabNewLine 
\Strut Primary Mirror Focal Ratio   & f/3.8			  \TabNewLine 
\Strut Effective Focal Length	    &  342.900 cm		  \TabNewLine 
\Strut System Focal Ratio	    & f/9.0			  \TabNewLine 
\Strut Obscuration Ratio	    &	 0.41			  \TabNewLine 
\Strut Field of View		    &	40\arcmin\ 		  \TabNewLine 
\Strut Plate Scale		    &	56.8\arcsec/mm  	  \TabNewLine 
\Strut Angular Resolution	    &	 2.5\arcsec\ 		 \TabNewLine 
\Strut Wavelength Range 	    & 1200-3200 \AA		  \TabNewLine 
\Strut Detectors		    & \raggedright Image intensifiers with 
				     \csi\ and \cste\ photocathodes
				       70-mm IIa-O film  	  \TabNewLine 
\Strut Magnitude limit  	    & \raggedright V=25 for S/N=10 observation
				     (30~min exp of 30000K star)  \TabNewLine 
\DTabLine
\end{tabular}
\end{centertable}
\begin{centertable}
\parcaption{3.8in}{\Astro{1}\ \UIT\ PSF Sizes.  \Astro{2}\ are 
similar.\label{uitpsfdesc}}
\begin{tabular}{lcc}
\Strut\TabNewLine
\DTabLine
\Strut  		& NUV FWHM & FUV FWHM \TabNewLine
			& arcsec   & arcsec   \TabNewLine \TabLine
\multicolumn{3}{l}{\Strut On-film values: (after deconvolution)}\TabNewLine 
Nominal pointing:	&	   &	      \TabNewLine
Whole image:		&      2.4 &	  2.9 \TabNewLine
R$<$16\arcmin\ 		&      2.2 &	  2.5 \TabNewLine \TabLine
\multicolumn{3}{l}{\Strut Digitized values: (before deconvolution)}\TabNewLine
Nominal pointing:	&	   &	      \TabNewLine
Whole image:		&      2.9 &	  3.4 \TabNewLine
R$<$16\arcmin\ 		&      2.7 &	  3.0 \TabNewLine \TabLine
\multicolumn{3}{l}{\Strut Off-nominal pointing:}\TabNewLine
Whole image:		&      3.3 &	  3.9 \TabNewLine
R$<$16\arcmin:  	&      3.1 &	  3.5 \TabNewLine \DTabLine
\end{tabular}
\end{centertable}
\begin{minipagetable}{3.75in}
\centering{\parcaption{1.6in}{\UIT\ Filters\label{uitfilters}}}
\begin{tabular}{ccrrr}
\Strut\TabNewLine
\DTabLine
\Strut Filter & Substrate & \multicolumn{1}{c}{$\lambda_{eff}$} & 
\multicolumn{1}{c}{Peak $\lambda$} & 
\multicolumn{1}{c}{$\Delta\lambda$\footnote{Area under sensitivity
curve of scale height 1.}} 
\TabNewLine
& & \multicolumn{1}{c}{\AA} & \multicolumn{1}{c}{\AA} & 
\multicolumn{1}{c}{\AA} \TabNewLine \TabLine
\Strut A1 & fused \siotwo    & 2488 & 2763	& 1147 \TabNewLine
\Strut A2 & fused \siotwo    & 1892 & 1853	&  412 \TabNewLine
\Strut A3 & fused \siotwo    & 1964 & 1899	&  173 \TabNewLine
\Strut A4 & fused \siotwo    & 2205 & 2184	&  244 \TabNewLine
\Strut A5 & fused \siotwo    & 2558 & 2508	&  456 \TabNewLine
\Strut A6 &  \caftwo	     & \multicolumn{1}{c}{\threedash} & 
\multicolumn{1}{c}{\threedash} & \multicolumn{1}{c}{\threedash} \TabNewLine 
\TabLine
\Strut B1 &  \srftwo	     & 1521 & 1443	&  354 \TabNewLine
\Strut B2 &  \caftwo	     & 1359 & 1266	&  160 \TabNewLine
\Strut B3 &  \caftwo	     & 1445 & 1385	&  256 \TabNewLine
\Strut B4 &  crystal \siotwo & 1585 & 1523,1676 &  129 \TabNewLine
\Strut B5 &  crystal \siotwo & 1615 & 1518	&  225 \TabNewLine
\Strut B6 &  \mgftwo	     & 1496 & 1477	&  404 \TabNewLine
\DTabLine
\end{tabular}
\end{minipagetable}
\begin{centertable}
\parcaption{3.25in}{Location of the Flat Field Defects.\label{ffdefects}}
\begin{tabular}{rrrrrrrrr}
\Strut\TabNewLine
\DTabLine
 \Strut & \multicolumn{1}{c}{X} & \multicolumn{1}{c}{Y} & & 
 \multicolumn{1}{c}{X} & 
 \multicolumn{1}{c}{Y} & & \multicolumn{1}{c}{X} & \multicolumn{1}{c}{Y} 
 \TabNewLine
\TabLine
 \Strut 1 &  710 &  244 &   7 &  787 &  216 & 13 & 1813 & 1142 \TabNewLine
	2 &  505 &  408 &   8 & 1195 &  867 & 14 &  226 & 1517 \TabNewLine
	3 &  817 &  522 &   9 & 1687 &  825 & 15 &  787 & 1752 \TabNewLine
	4 &  926 &  833 &  10 & 1145 &  967 & 16 &  737 & 1022 \TabNewLine
	5 &  469 &  683 &  11 & 1049 & 1727 & 17 &  203 & 1387 \TabNewLine
	6 &  866 &  381 &  12 & 1597 & 1634 &	 &	&      \TabNewLine
\DTabLine
\end{tabular}
\end{centertable}
\begin{tableletters}
\begin{centertable}
\parcaption{3.5in}{Astro 1 NUV Camera Image Distortion Model Coefficients
\label{nuvimgdistmodelA}}
\begin{tabular}{lr@{$\times 10$}lr@{$\times 10$}lr@{$\times 10$}lr@{$\times
10$}l}
\Strut\TabNewLine
\DTabLine
 & \multicolumn{4}{c}{Raw-to-Ideal} & \multicolumn{4}{c}{Ideal-to-Raw}
\TabNewLine
 & \multicolumn{2}{c}{$x'$} & \multicolumn{2}{c}{$y'$}
 & \multicolumn{2}{c}{$x'$} & \multicolumn{2}{c}{$y'$} \TabNewLine
\TabLine
$a_{0}$ & $ 5.40$ & $^{+01}$ & $ 4.34$ & $^{+01}$ & $-5.67$ & $^{+01}$ & $-4.55$
& $^{+01}$ \TabNewLine
$a_{1}$ & $ 9.25$ & $^{-01}$ & $-6.41$ & $^{-02}$ & $ 1.08$ & $^{+00}$ & $ 6.72$
& $^{-02}$ \TabNewLine
$a_{2}$ & $-8.72$ & $^{-02}$ & $ 9.27$ & $^{-01}$ & $ 9.13$ & $^{-02}$ & $ 1.08$
& $^{+00}$ \TabNewLine
$a_{3}$ & $ 4.86$ & $^{-05}$ & $ 3.25$ & $^{-05}$ & $-5.10$ & $^{-05}$ & $-3.38$
& $^{-05}$ \TabNewLine
$a_{4}$ & $ 6.52$ & $^{-05}$ & $ 6.03$ & $^{-05}$ & $-6.88$ & $^{-05}$ & $-6.35$
& $^{-05}$ \TabNewLine
$a_{5}$ & $ 4.22$ & $^{-05}$ & $ 4.70$ & $^{-05}$ & $-4.39$ & $^{-05}$ & $-4.89$
& $^{-05}$ \TabNewLine
$a_{6}$ & $-1.72$ & $^{-08}$ & $-5.28$ & $^{-10}$ & $ 1.79$ & $^{-08}$ & $ 4.55$
& $^{-10}$ \TabNewLine
$a_{7}$ & $ 3.09$ & $^{-09}$ & $-2.71$ & $^{-08}$ & $-3.18$ & $^{-09}$ & $ 2.85$
& $^{-08}$ \TabNewLine
$a_{8}$ & $-3.62$ & $^{-08}$ & $-5.14$ & $^{-09}$ & $ 3.80$ & $^{-08}$ & $ 5.38$
& $^{-09}$ \TabNewLine
$a_{9}$ & $ 1.93$ & $^{-09}$ & $-1.35$ & $^{-08}$ & $-2.13$ & $^{-09}$ & $ 1.40$
& $^{-08}$ \TabNewLine
\DTabLine
\end{tabular}
\end{centertable}
\begin{centertable}
\parcaption{3.5in}{Astro 1 FUV Camera Image Distortion Model Coefficients
\label{fuvimgdistmodelA}}
\begin{tabular}{lr@{$\times 10$}lr@{$\times 10$}lr@{$\times 10$}lr@{$\times
10$}l}
\Strut\TabNewLine
\DTabLine
 & \multicolumn{4}{c}{Raw-to-Ideal} & \multicolumn{4}{c}{Ideal-to-Raw}
\TabNewLine
 & \multicolumn{2}{c}{$x'$} & \multicolumn{2}{c}{$y'$}
 & \multicolumn{2}{c}{$x'$} & \multicolumn{2}{c}{$y'$} \TabNewLine
\TabLine
$a_{0}$ & $ 1.65$ & $^{+01}$ & $ 8.66$ & $^{+01}$ & $-1.63$ & $^{+01}$ & $-9.30$
& $^{+01}$ \TabNewLine
$a_{1}$ & $ 9.10$ & $^{-01}$ & $-1.11$ & $^{-01}$ & $ 1.09$ & $^{+00}$ & $ 1.21$
& $^{-01}$ \TabNewLine
$a_{2}$ & $ 1.96$ & $^{-02}$ & $ 8.43$ & $^{-01}$ & $-2.41$ & $^{-02}$ & $ 1.17$
& $^{+00}$ \TabNewLine
$a_{3}$ & $ 7.93$ & $^{-05}$ & $ 6.41$ & $^{-05}$ & $-8.40$ & $^{-05}$ & $-7.04$
& $^{-05}$ \TabNewLine
$a_{4}$ & $ 3.80$ & $^{-05}$ & $ 9.04$ & $^{-05}$ & $-3.96$ & $^{-05}$ & $-9.67$
& $^{-05}$ \TabNewLine
$a_{5}$ & $-3.10$ & $^{-05}$ & $ 1.12$ & $^{-04}$ & $ 3.62$ & $^{-05}$ & $-1.21$
& $^{-04}$ \TabNewLine
$a_{6}$ & $-3.27$ & $^{-08}$ & $-9.26$ & $^{-09}$ & $ 3.53$ & $^{-08}$ & $ 1.04$
& $^{-08}$ \TabNewLine
$a_{7}$ & $ 1.91$ & $^{-08}$ & $-3.56$ & $^{-08}$ & $-2.18$ & $^{-08}$ & $ 3.83$
& $^{-08}$ \TabNewLine
$a_{8}$ & $-3.74$ & $^{-08}$ & $-6.86$ & $^{-09}$ & $ 4.09$ & $^{-08}$ & $ 7.22$
& $^{-09}$ \TabNewLine
$a_{9}$ & $ 1.94$ & $^{-08}$ & $-3.37$ & $^{-08}$ & $-2.19$ & $^{-08}$ & $ 3.66$
& $^{-08}$ \TabNewLine
\DTabLine
\end{tabular}
\end{centertable}
\begin{centertable}
\parcaption{3.5in}{Astro 2 FUV Camera Image Distortion Model Coefficients
\label{fuvimgdistmodelB}}
\begin{tabular}{lr@{$\times 10$}lr@{$\times 10$}lr@{$\times 10$}lr@{$\times
10$}l}
\Strut\TabNewLine
\DTabLine
 & \multicolumn{4}{c}{Raw-to-Ideal} & \multicolumn{4}{c}{Ideal-to-Raw}
\TabNewLine
 & \multicolumn{2}{c}{$x'$} & \multicolumn{2}{c}{$y'$}
 & \multicolumn{2}{c}{$x'$} & \multicolumn{2}{c}{$y'$} \TabNewLine
\TabLine
$a_{0}$ & $ 1.87$ & $^{+01}$ & $ 9.30$ & $^{+01}$ & $-1.81$ & $^{+01}$ & $-1.01$
& $^{+02}$ \TabNewLine
$a_{1}$ & $ 9.07$ & $^{-01}$ & $-1.19$ & $^{-01}$ & $ 1.10$ & $^{+00}$ & $ 1.30$
& $^{-01}$ \TabNewLine
$a_{2}$ & $ 1.94$ & $^{-02}$ & $ 8.29$ & $^{-01}$ & $-2.50$ & $^{-02}$ & $ 1.19$
& $^{+00}$ \TabNewLine
$a_{3}$ & $ 8.13$ & $^{-05}$ & $ 6.76$ & $^{-05}$ & $-8.61$ & $^{-05}$ & $-7.48$
& $^{-05}$ \TabNewLine
$a_{4}$ & $ 3.72$ & $^{-05}$ & $ 9.90$ & $^{-05}$ & $-3.81$ & $^{-05}$ & $-1.07$
& $^{-04}$ \TabNewLine
$a_{5}$ & $-3.17$ & $^{-05}$ & $ 1.24$ & $^{-04}$ & $ 3.77$ & $^{-05}$ & $-1.36$
& $^{-04}$ \TabNewLine
$a_{6}$ & $-3.32$ & $^{-08}$ & $-9.96$ & $^{-09}$ & $ 3.57$ & $^{-08}$ & $ 1.13$
& $^{-08}$ \TabNewLine
$a_{7}$ & $ 1.93$ & $^{-08}$ & $-3.70$ & $^{-08}$ & $-2.21$ & $^{-08}$ & $ 4.02$
& $^{-08}$ \TabNewLine
$a_{8}$ & $-3.76$ & $^{-08}$ & $-1.01$ & $^{-08}$ & $ 4.09$ & $^{-08}$ & $ 1.10$
& $^{-08}$ \TabNewLine
$a_{9}$ & $ 2.01$ & $^{-08}$ & $-3.69$ & $^{-08}$ & $-2.29$ & $^{-08}$ & $ 4.03$
& $^{-08}$ \TabNewLine
\DTabLine
\end{tabular}
\end{centertable}
\end{tableletters}
\begin{centertable}
\parcaption{3in}{\UIT\ Measurement Uncertainties\label{uituncert}}
\begin{tabular}{llp{1.5in}p{1.5in}}
\Strut\TabNewLine
\DTabLine
\multicolumn{1}{c}{\Strut Measurement} & \multicolumn{1}{c}{Uncertainty}
 & \multicolumn{1}{c}{Origin} & \multicolumn{1}{c}{Caveats}\TabNewLine
\multicolumn{1}{c}{} & \multicolumn{1}{c}{(Sigma)} 
& \multicolumn{1}{c}{} & \multicolumn{1}{c}{}\TabNewLine \TabLine
\Strut Exposure Level & 1 E-unit & Film \& 
\raggedright microdensitometer fog level uncertainty &
\raggedright Important for low surface brightness\TabNewLine
\Strut Absolute calibration  &  15\%  & 
\raggedright \IUE\ uncertainty \&\ \UIT\ ties to it & 
\TabNewLine
\Strut Stellar photometry &  0.1 mag & Integrated S/N; ap. \& sky corrections & 
\raggedright well-exposed stars (peaks 20$<$E$<$500) \TabNewLine 
Surface photometry & 1-pixel: 15\%  & Film \& char. curve &
\raggedright (30$<$E$<$300, local spatial varns. add 3\%) \TabNewLine				      
& 25-pixel: 5\%  & Film \& char. curve & \TabNewLine
\Strut Position 	  & 3\arcsec\   & Image distortions & 
\raggedright R(field)$<$16\arcmin\ after correction \TabNewLine \DTabLine
\end{tabular}
\end{centertable}
\begin{centertable}
\parcaption{3.5in}{Summary of \UIT\ Image Data Products\label{dataprodsumm}}
\begin{tabular}{lccl}
\Strut\TabNewLine
\DTabLine
\Strut  & Image &	     & \multicolumn{1}{c}{Sample}\TabNewLine
Product & Size  & Astrometry & \multicolumn{1}{c}{Filename}\TabNewLine
\TabLine\Strut
\Astro{1}\ Raw Density	      & $2048 \times 2048$ & No  & fuv0123d.fit\TabNewLine
\Astro{2}\ Raw Density	      & $4096 \times 4096$ & No  & fuv2123d.fit\TabNewLine
Linearized, Flat fielded      & $2048 \times 2048$ & No  & fuv0123e.fit\TabNewLine
Linearized, Flat fielded      & $2048 \times 2048$ & Yes & fuv0123f.fit\TabNewLine
Linearized, Flat fielded      &  $512 \times 512$  & Yes & fuv0123fc.fit\TabNewLine
Undistorted, Rotated North-up & $2048 \times 2048$ & Yes & fuv0123g.fit\TabNewLine
Undistorted, Rotated North-up &  $512 \times 512$  & Yes & fuv0123gc.fit\TabNewLine  
\DTabLine
\end{tabular}
\end{centertable}
%
%
%
%
\clearpage
\section*{Figures}

\begin{centerfigure}
\centerepsf{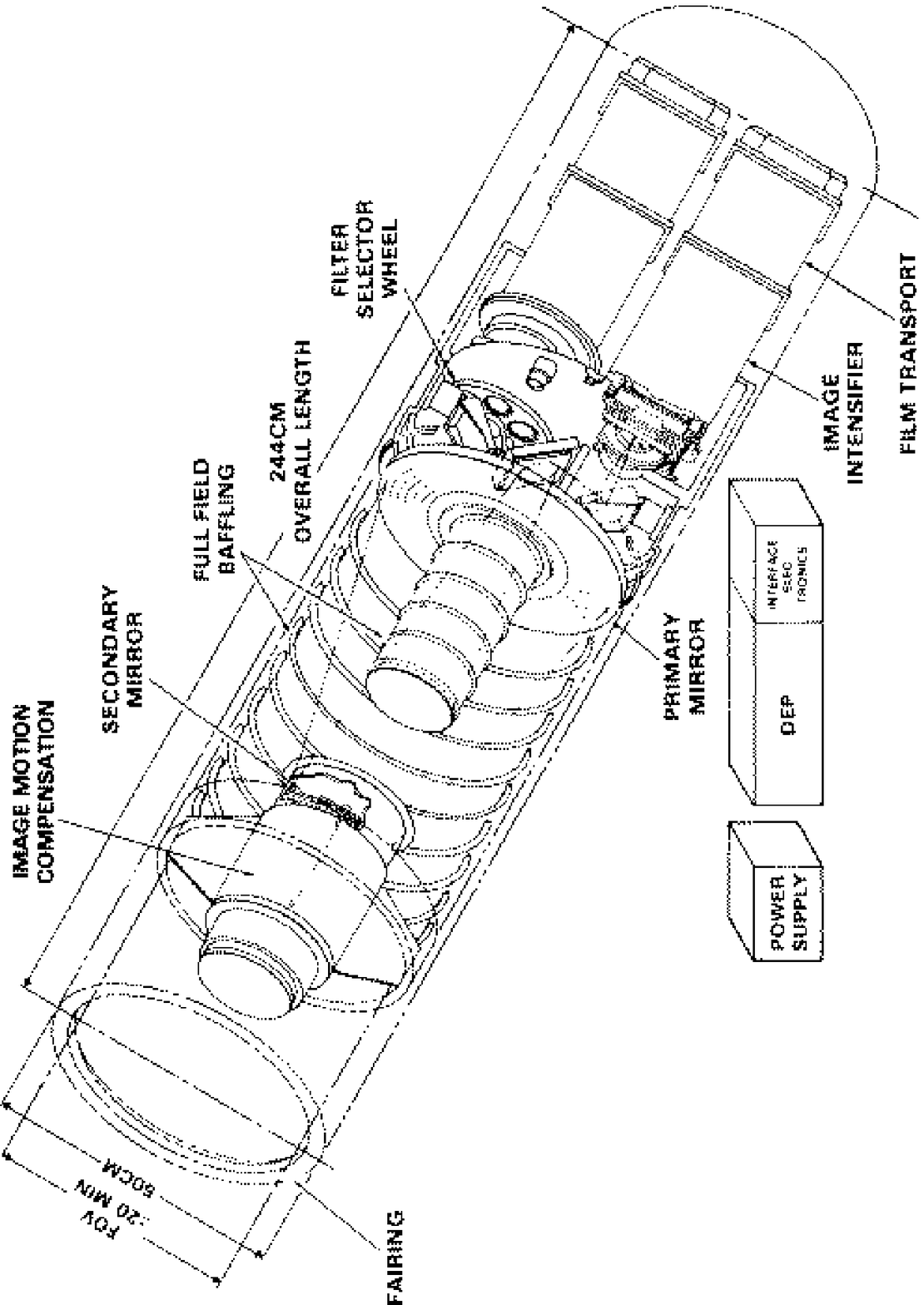}
\caption{A cut-away view of the \UIT.\label{uitcutaway}}
\end{centerfigure}
\begin{centerfigure}
\centerepsf{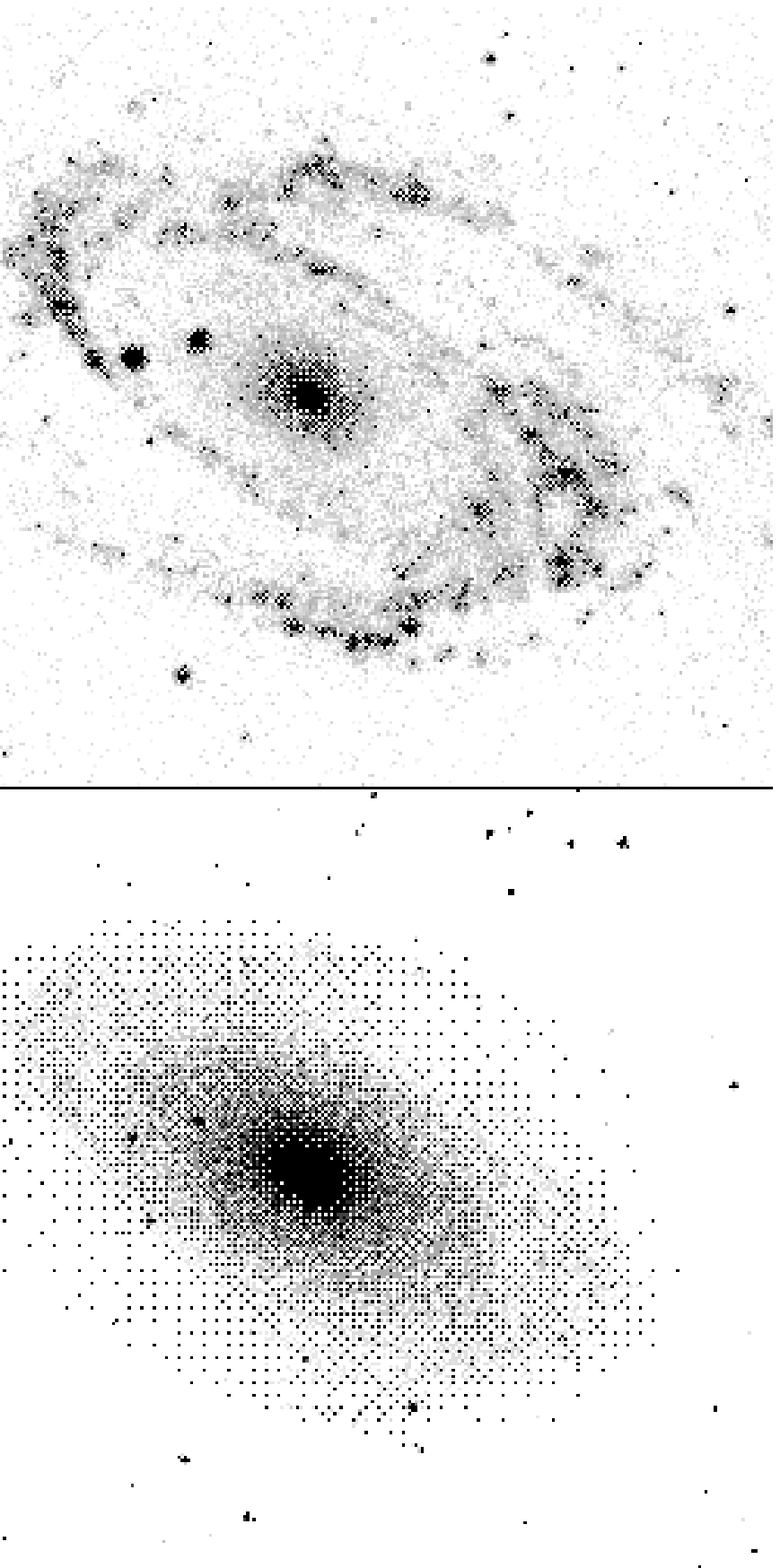}
\caption{An \Astro{1}\ \UIT\ 639-second A1 image
($\lambda_{eff}=2490$\AA) of the spiral galaxy M81 (left) and a ground-based
visible-band image made at Kitt Peak National Observatory (right). The UV image
accentuates the UV-bright Population I component in the spiral arms, as
well as the hot old-population stars in  the nuclear bulge. The weakness of
UV emission from the general disk  population in this early-type spiral is
typical (\protect\cite{jhillc}); later type  spirals show stronger UV flux 
from the disk (\protect\cite{cornetta}).\label{mplate}}
\end{centerfigure}
\begin{centerfigure}
\centerepsf{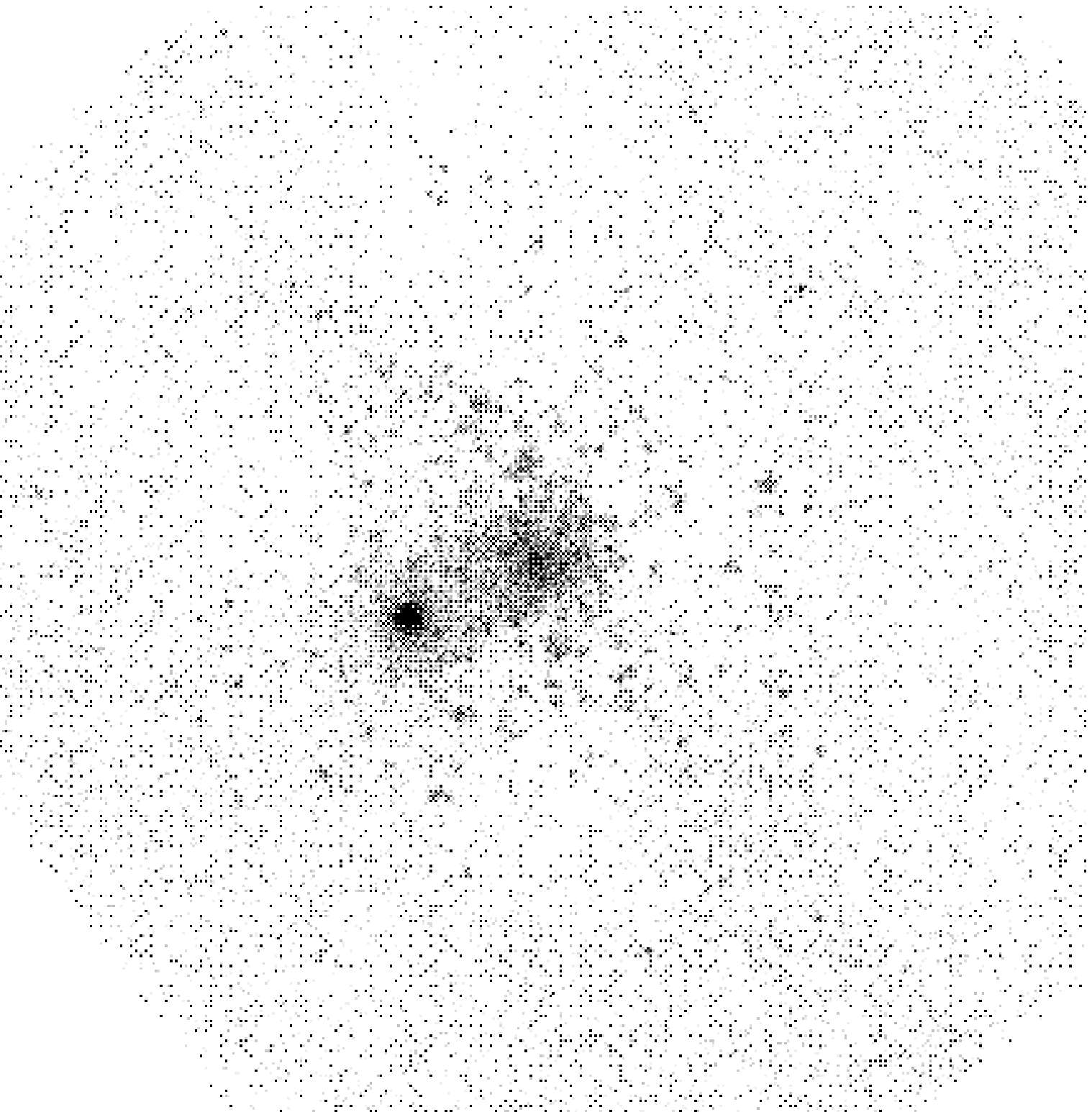}
\caption{An \Astro{2}\ \UIT\ 781 second exposure of the globular cluster
NGC~6752 made in the B5 filter ($\lambda_{eff}=1620$\AA).
The rejection of visible-band light suppresses the cluster's $\sim 100,000$
main sequence stars, leaving only the 355 hot, UV-bright horizontal branch 
stars in the field of view.  The \UIT\ image resolves hot
stars in the cluster core, and the 40 arcminute field of view encompasses
the entire cluster (\protect\cite{landsmand}). The overexposed bright object
is a foreground star.\label{ngcplate}}
\end{centerfigure}
\begin{centerfigure}
\centerepsf{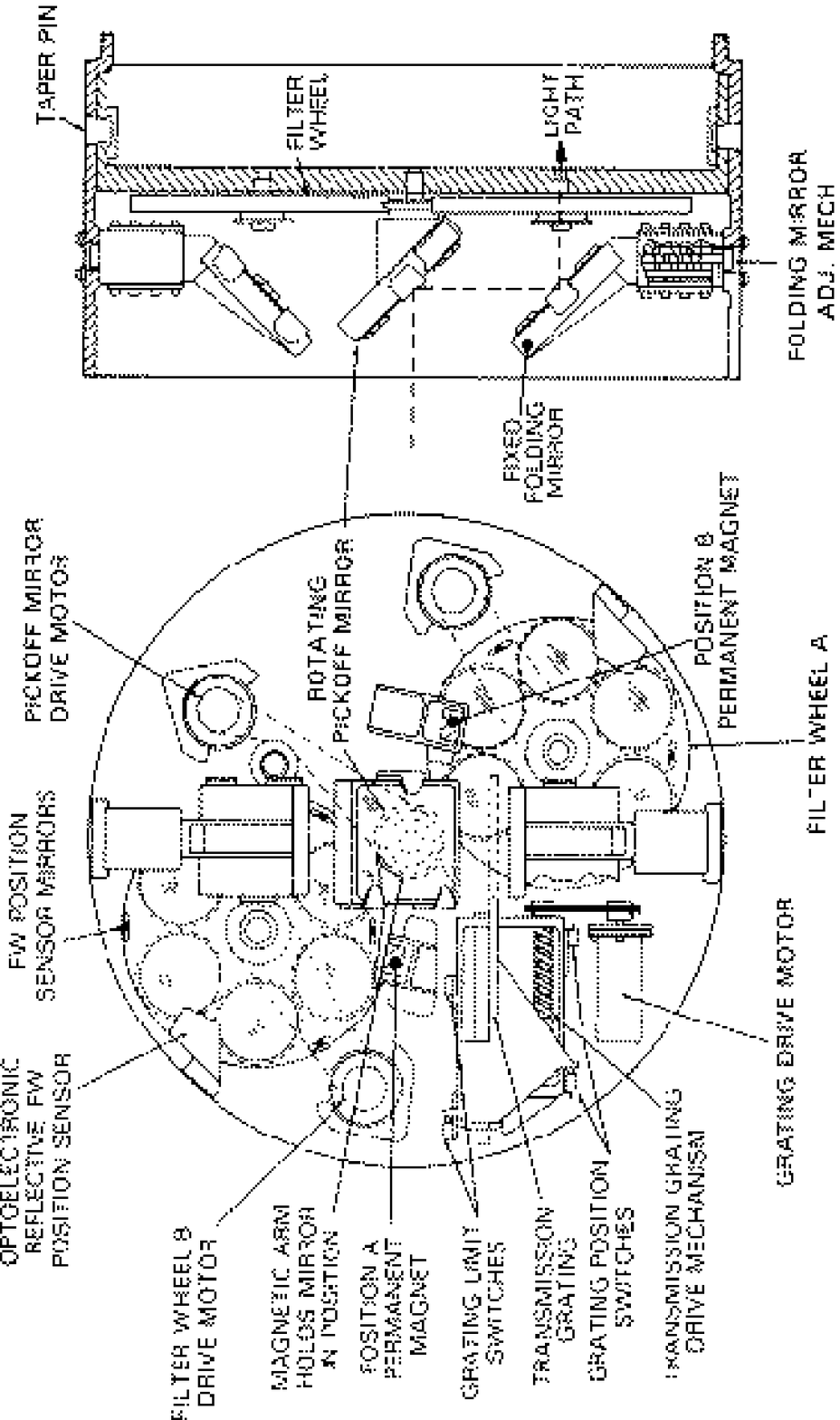}
\caption{A view of the \UIT\ suboptical assembly.\label{suboptical}}
\end{centerfigure}
\begin{figureletters}
\begin{centerfigure}
\centerepsf{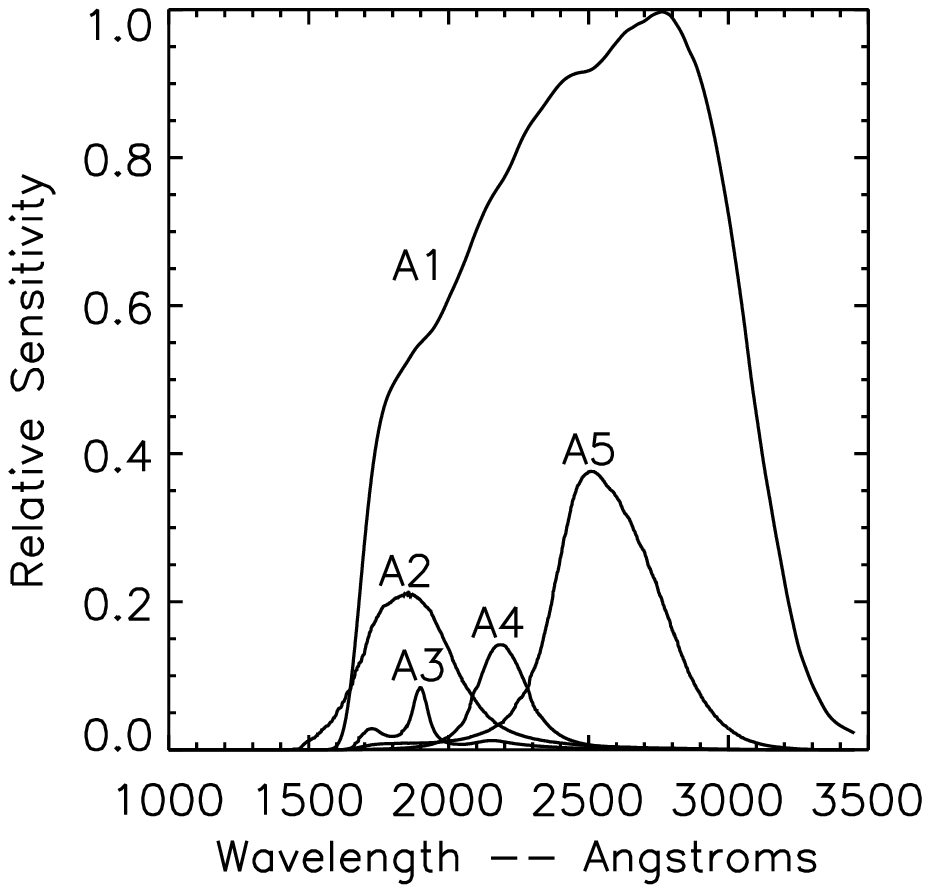}
\caption{The near-UV instrument response curves for each filter.
\label{nuvcurves}}
\end{centerfigure}
\clearpage
\begin{centerfigure}
\centerepsf{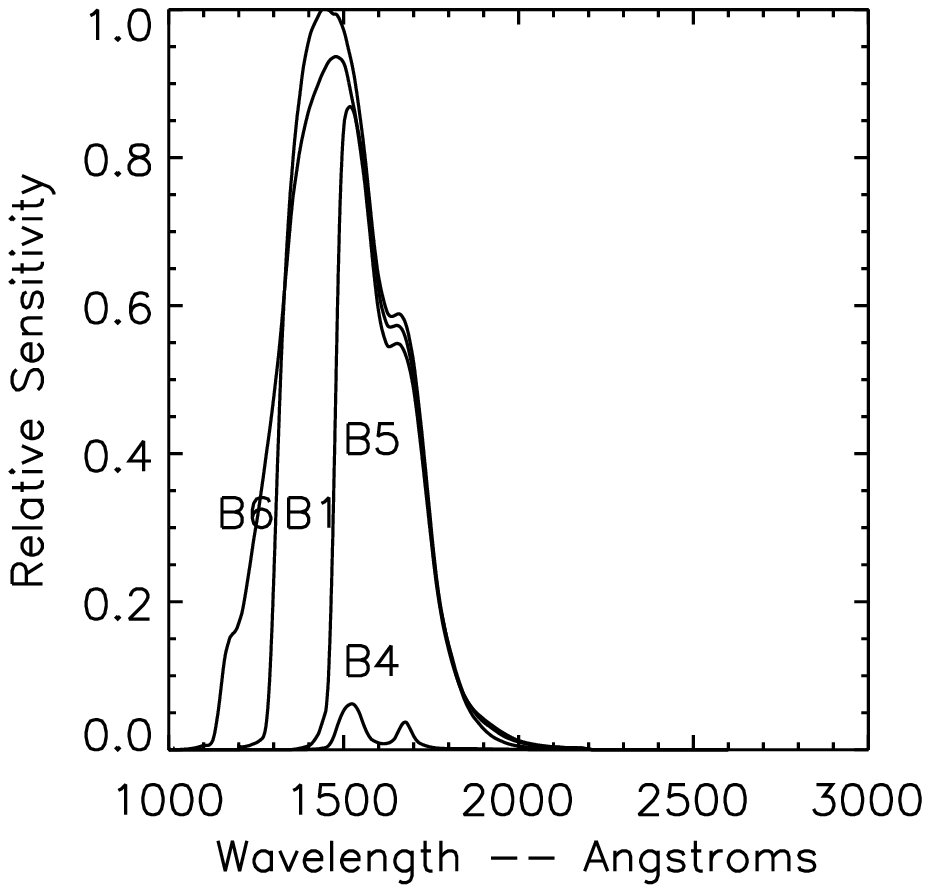}
\caption{The far-UV instrument response curves for each filter.
\label{fuvcurves}}
\end{centerfigure}
\end{figureletters}
\begin{centerfigure}
\centerepsf{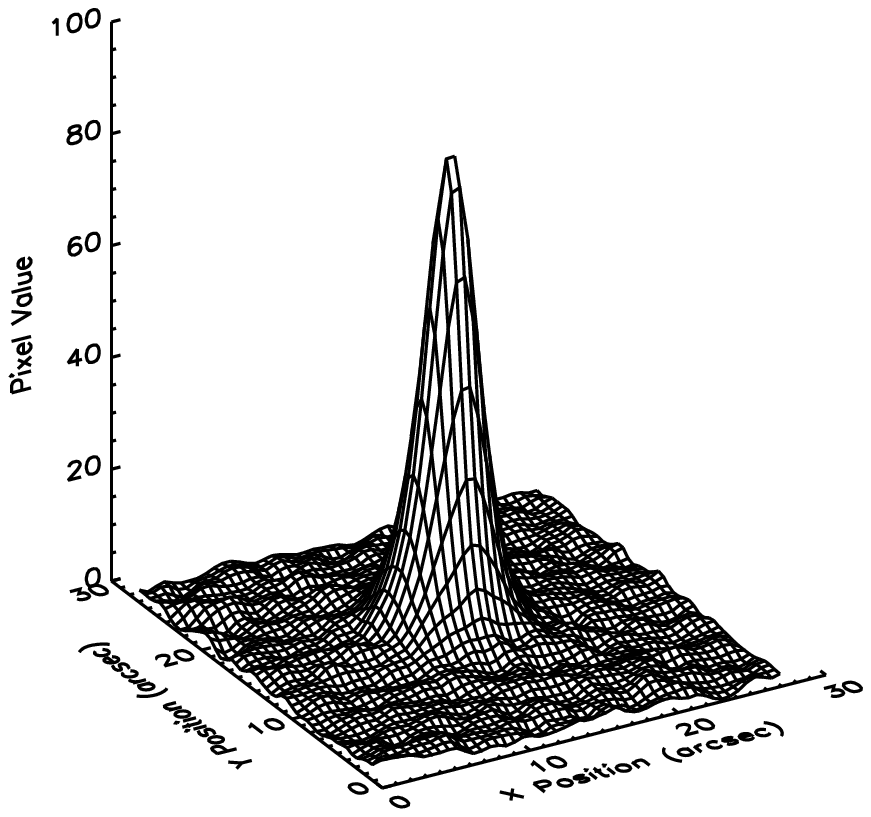}
\caption{The \UIT\ FUV PSF as measured from star images on the \Astro{1}\ 
frame FUV0091.\label{psfsurf}}
\end{centerfigure}
\begin{centerfigure}
\centerepsf{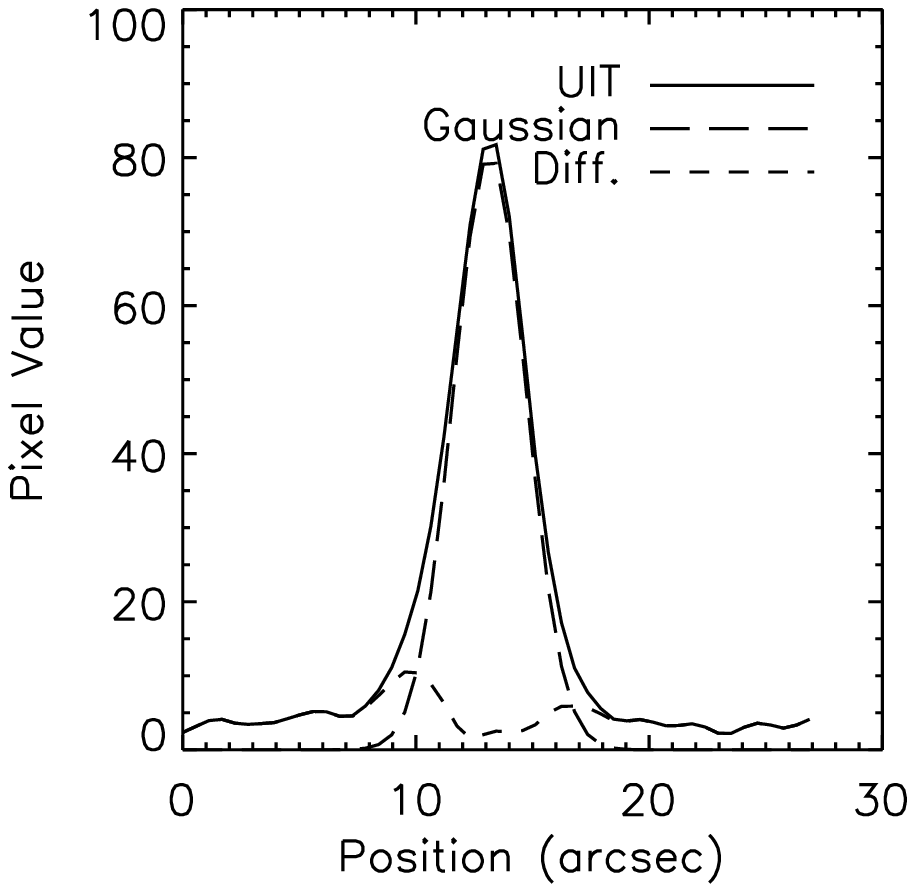}
\caption{The \UIT\ FUV PSF: as measured (upper solid curve); as fit by a 
Gaussian
function (lower solid curve); and the residual difference between the 
two.\label{diffpsf}}
\end{centerfigure}
\begin{centerfigure}
\centerepsf{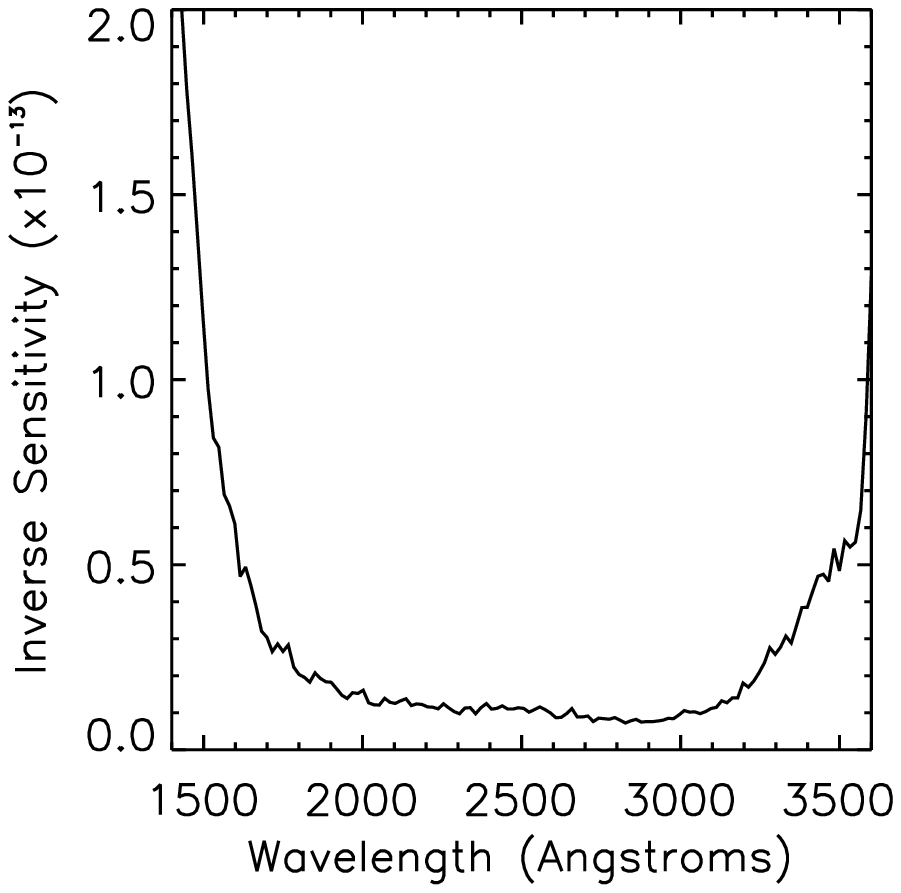}
\caption{Sensitivity curve as a function of wavelength for the \UIT\
grating, derived from in-flight observations of the white dwarf calibrator
GB191B2B.  An \IUE\ spectrum of this star was compared to the \UIT\
spectrum extracted from a grating image.  This curve is expressed in units
of \eunitps\ / \surfbr.\label{gratecurv}}
\end{centerfigure}
\begin{centerfigure}
\centerepsf{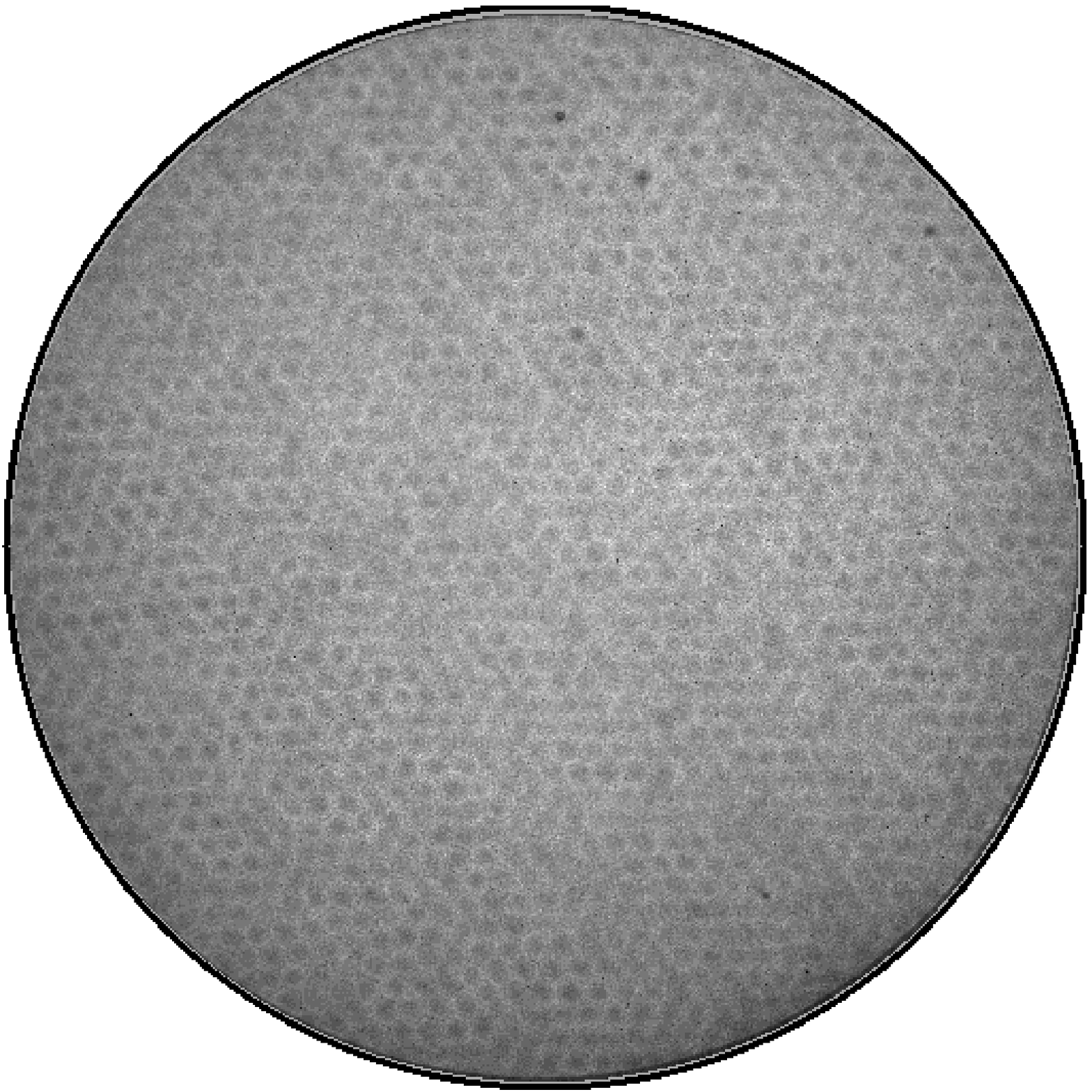}
\caption{The FLIGHT22 composite flat field for filter ``B5'', median-summed from
8 individual laboratory flat field exposures.  The honeycomb pattern is caused
by a similar pattern on the fiber optics which couple the image intensifier
output to the film.  As scaled to flatten science images, the darkest pixels 
in this image have values $\sim 170$ E-units, and the lightest pixels have 
values $\sim 270$ E-units.\label{flatfield}}
\end{centerfigure}
\begin{centerfigure}
\centerepsf{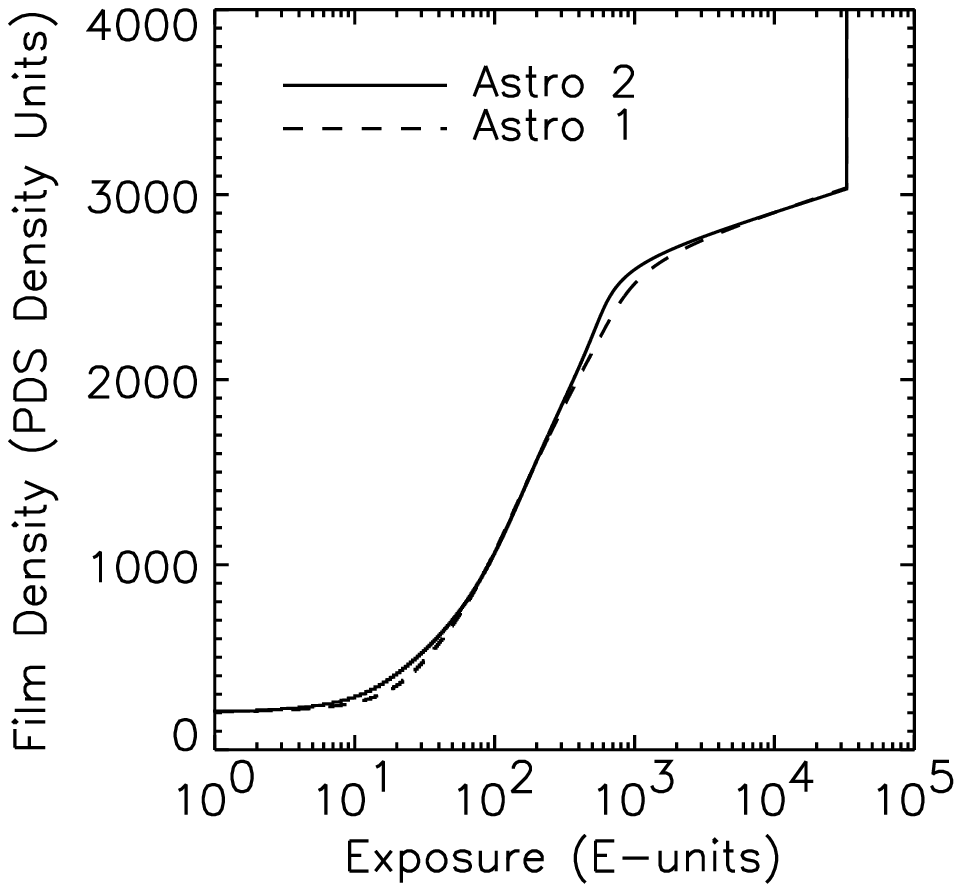}
\caption{The \Astro{1}\ and \Astro{2}\ characteristic curves.
The \Astro{1}\ curve was used for FLIGHT15 and FLIGHT21 processing.  The
\Astro{2}\ curve was used for FLIGHT22 processing.\label{charcurvplot}}
\end{centerfigure}
\begin{centerfigure}
\centerepsf{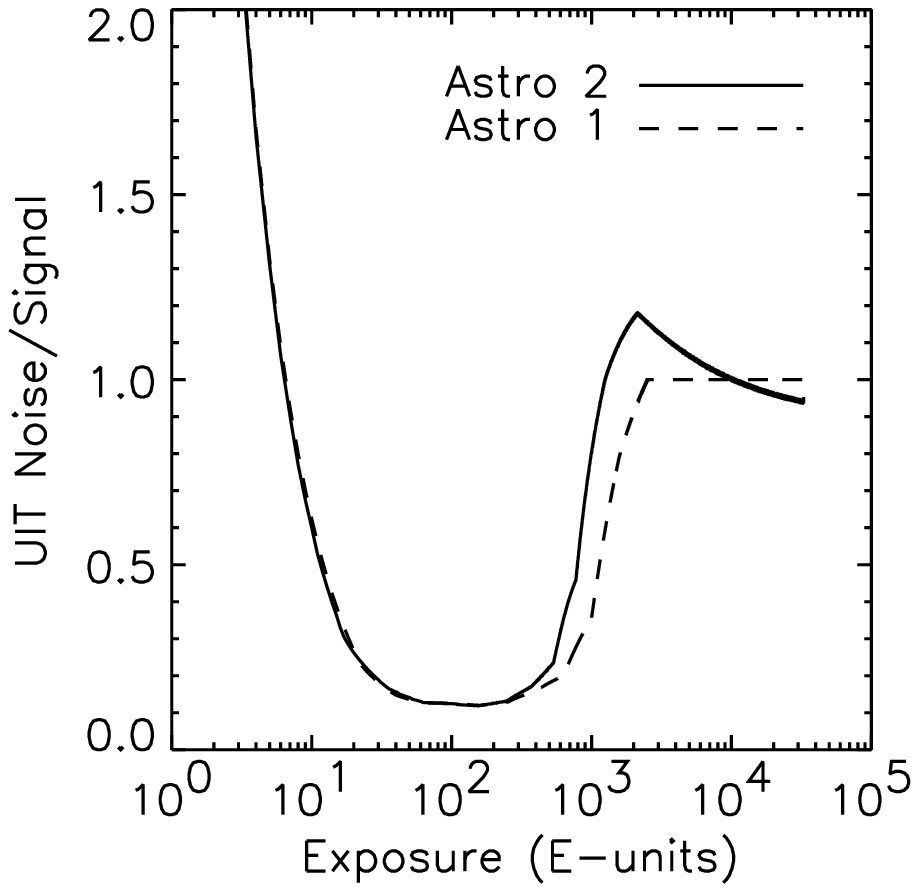}
\caption{The \Astro{1}\ (FLIGHT15) and \Astro{2}\ (FLIGHT22)
noise-to-signal curves.  \label{noisecurvplot}}
\end{centerfigure}
\begin{centerfigure}
\centerepsf{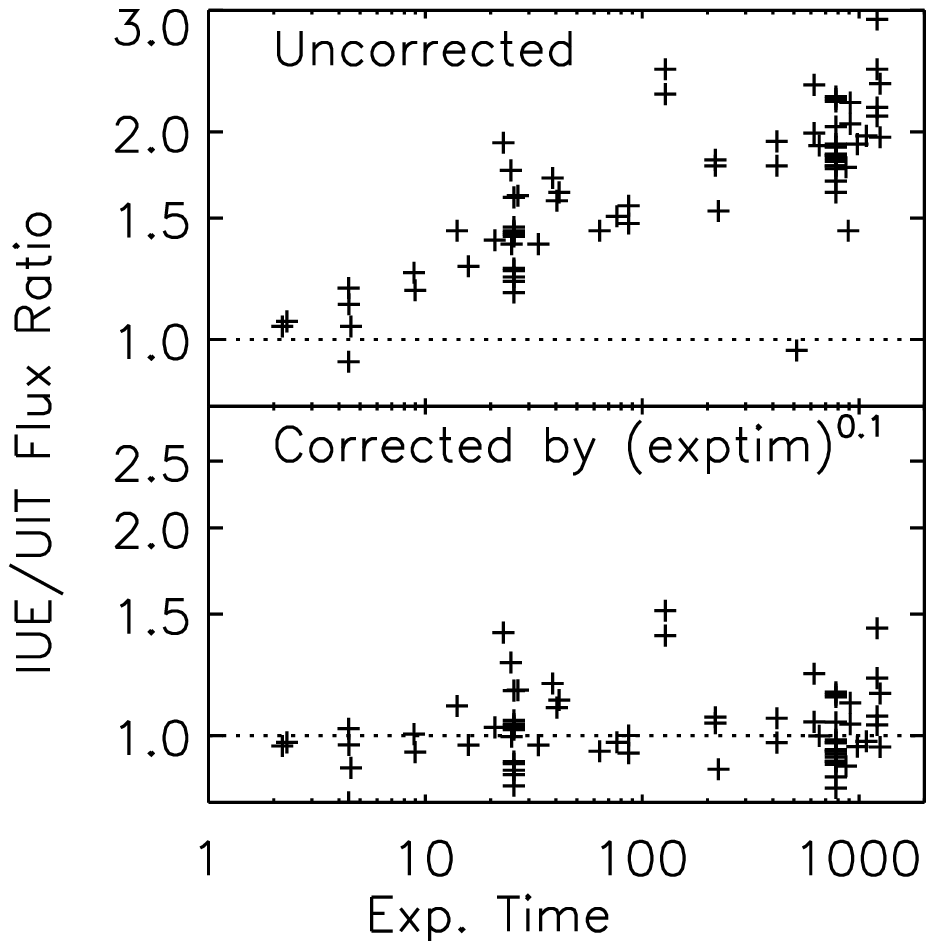}
\caption{ The ratio of the \IUE\ to the \Astro{2}\ \UIT\ flux is shown for
75 measurements of 59 stars.  In the top panel, this flux ratio is seen to
be a strong function of the exposure time of the \UIT\ image.  The origin
of this dependency is not understood, but it is reminiscent of reciprocity
failure. Multiplying the \UIT\ flux by the exposure time raised to the 0.1
power removes most of this dependency, and it allows the use of a single
basic calibration constant.\label{abscalfig}}
\end{centerfigure}
\end{document}